\documentclass[12pt,letter,superscriptaddress,prl,aps]{revtex4-1}
\pdfoutput=1
\usepackage[utf8x]{inputenc}
\usepackage{amsmath,amssymb}
\usepackage{graphicx}
\usepackage{dcolumn}
\usepackage{bm}
\usepackage{subfigure}
\usepackage{enumerate}
\usepackage{multirow}
\usepackage{tabularx}
\usepackage{threeparttable}
\usepackage{epstopdf}
\usepackage{paralist}
\usepackage{lineno}
\usepackage{color}
\usepackage{soul}
\sethlcolor{yellow}

\newcommand{\ket}[1]{\left| #1 \right\rangle}

\newcommand{\COMMENT}[1]{}
\newcommand{\textapprox}{{\raise.17ex\hbox{$\scriptstyle\mathtt{\sim}$}}}
\usepackage{accents}
\newcommand*{\dt}[1]{%
  \accentset{\mbox{\large\bfseries .}}{#1}}

\begin{document}

\title{A multi $k$-point nonadiabatic molecular dynamics for periodic systems}

\author{Fan Zheng}
 \affiliation{School of physical science and technology, ShanghaiTech University, Shanghai 201210, China. }
 \email{zhengfan@shanghaitech.edu.cn}
\author{Lin-wang Wang}
 \affiliation{State key laboratory of superlattices and microstructures, Institute of semiconductors, Chinese Academy of Science, Beijing 100083, China.}
 \email{lwwang@semi.ac.cn}

\begin{abstract}

With the rapid development of ultra-fast experimental techniques used for carrier dynamics in solid-state systems, a microscopic understanding of the related phenomena, particularly a first-principle calculation is highly desirable. Non-adiabatic molecular dynamics (NAMD) offers a real-time direct simulation of the carrier transfer or carrier thermalization. However, when applied to a periodic supercell, due to the $\Gamma$-point phonon movement during the molecular dynamics, there is no supercell electronic $k$-point crossing during the NAMD simulation. This often leads to a significant underestimation of the transition rate due to significant energy gaps in the single supercell $k$-point band structure.  In this work, based on the surface hopping scheme used for NAMD, we propose a practical method to enable the cross-$k$ transition for a periodic system. We demonstrate our formalism by showing that the hot electron thermalization process by the multi $k$-point NAMD in a small supercell is equivalent to such simulation in a large supercell with single $\Gamma$ $k$-point. The hot carrier thermalization process in the bulk silicon is also carried out and compared with the recent ultra-fast experiments with excellent agreements.

\end{abstract}

\maketitle
\newpage

The ultrafast carrier dynamics in solid states as illustrated by the transport and thermalization processes plays a critical role in various areas, such as in energy materials~\cite{Heeger14p10,Shi18p879}, optoelectronic devices~\cite{Jhalani17p5012,Gupta92p2464,BowersII09p5730}, sensors~\cite{Born16p2475}, and charge/spin transport in quantum materials~\cite{Dawlaty08p042116}. Benefit from the recent rapid improvements of spatial and temporal resolutions of the ultrafast experimental techniques, we are now able to trace the very detail and fundamental dynamical movements of electrons, holes, and ions~\cite{Rossi02p895}. However, the development of theoretical tools, in particular first-principle methods, to study the ultrafast carrier transport process is lagging behind. The large size of the systems needed in order to reveal the underlying physics in applications such as solar cells, light-emitting diodes, catalytical surfaces, and quantum devices brings tremendous challenges in \textit{ab initio} simulations. Meanwhile, the complex nature of physical processes associated with different scattering mechanisms including electron-electron, electron-phonon, electron-hole, and carrier-defect goes beyond the equilibrium condition at which the most first-principle calculations are formulated~\cite{Rossi02p895}. One solution is to compute the various scatterings explicitly with first-principle perturbation methods~\cite{Giustino17p015003a,Bernardi14p257402}. This kind of methods provides an intuitive and systematic understanding of various couplings in carrier transport. Combined with Boltzmann transport equation, the dynamical pictures of carrier transfer or carrier thermalization could be revealed~\cite{Bernardi14p257402,Ponce20p036501,Jhalani17p5012,Sio19p,Bernardi15p5291}. However, for a complex system such as surface, interface, or nano-materials with thousands of energy bands and phonon modes, the perturbation calculation becomes cumbersome and difficult.

Non-adiabatic molecular dynamics (NAMD), originally developed in quantum chemistry field, has become an emerging way to study the ultrafast carrier dynamics in periodic solid states~\cite{Craig05p,Long11p19240,Long14p4343,Kang19p224303,Zheng19pe1411,Banerjee20p091102}. NAMD is carried out by directly evolving the wavefunctions of excited carriers following the time-dependent Schr\"{o}dinger's equation. Different types of NAMD are developed and implemented, including the Ehrenfest mean-field dynamics, surface hopping techniques, and multi-spawning methods~\cite{Crespo-Otero18p7026,Curchod18p3305}. Particularly, in solid states, by combining with the classical path approximation (CPA)~\cite{Akimov14p789}, the surface hopping NAMD simulation becomes a post-processing of ground-state \textit{ab initio} molecular dynamics (MD), which simplifies the implementations and reduces the computational cost greatly. This approximation is possible since the wavefunction evolution information is already encoded in the change of the eigen-states within MD. Here, the CPA utilizes the ions' dynamics in a fixed potential energy surface (usually ground state) and ignores the ``feedback" of excited carriers to the ionic motions. This approximation works well in solid-state materials when the carrier excitation is not heavy and the wavefunctions are extended. Based on the CPA, the NAMD simulations are carried out in various solid-state dynamical processes, such as charge thermalization~\cite{Kang19p224303,Zhou19p184701}, interfacial carrier transfer~\cite{Zheng19p6174,Wang21p2165,Zhang19p,Zheng17p6435,Long16p1996}, exciton transfer~\cite{Jiang21peabf3759}, surface chemical reactions~\cite{Zheng20p041115,Chu22peabo2675b}, and defect trapping~\cite{Chu20p6435,Zhang19p6151}. However, most NAMD simulations of solid states only allow a supercell with a single $\Gamma$-point wavevector sampling, even when the multi-$k$ points are used, the carrier can not transfer from one $k$ point to another. While this setup is generally fine for small nano materials, in the periodic bulk calculation, the affordable supercell size is usually not large enough. With a supercell MD trajectory, only the zone-center phonon modes of the supercell are represented, thus there is no electron transfer across different $k$ points due to phonon coupling. More importantly, in many cases, the single $k$-point electronic eigen energies of an affordable supercell has large artificial ``energy gaps", which causes artificial ``phonon bottleneck", and significantly slows down the thermalization process.

In this work, we propose a multi $k$-point first-principle NAMD approach based on the decoherence induced surface hopping (DISH) scheme to capture the electronic transitions between the eigen states across different $k$ points. By using the bulk silicon as an example, we demonstrate that the hot electron thermalization in a small supercell with a multi $k$-point sampling is consistent to its equivalent large supercell with a single-$\Gamma$-point simulation. This approach is further optimized to include the acoustic phonon modes induced by the deformation potential effect to assist the intra-band carrier transport. By comparing with the experimental hot electron thermalization in the bulk silicon, we show that this calculation yields an excellent agreement with the experiment. 

In an MD simulation using a supercell, there are only zone-center phonon modes assisting the inter-state transitions within a $k$ point. This is realized by computing the non-adiabatic coupling $V_{ij}(t)=\left\langle{\phi_i(t)}\middle|\frac{\partial{\phi_j(t)}}{\partial t}\right\rangle$, where $\ket{\phi_i(t)}$ is the time-dependent adiabatic states (i.e. eigen states in our NAMD scheme) with state index $i$. First, 

\begin{align}
    V_{ij}(t) & = \left< \phi_{i}(t) \middle| \frac{\partial \phi_{j}(t)}{\partial t} \right> \nonumber\\
              & = \sum_{\mathbf{R}} \left< \phi_{i}(t) \middle| \frac{\partial \phi_{j}(t)}{\partial \mathbf{R}} \right> \dt{\mathbf{R}} = \sum_{\mathbf{R}} \frac{1}{\epsilon_j - \epsilon_i}\left< \phi_{i}(t) \middle| \frac{\partial H}{\partial \mathbf{R}} \middle |\phi_{j}(t) \right> \dt{\mathbf{R}}
\end{align}

\noindent where $\mathbf{R}$ is the MD trajectory and $\epsilon_i$ is the eigen energy of state $i$. However, for a multi $k$-point scheme, a non-zone-center phonon mode with finite wavevector $\mathbf{q}$ will be needed to assist the carrier transition between different wavevectors $\mathbf{k}$ and $\mathbf{k'}$ (with $\mathbf{q}=\mathbf{k}-\mathbf{k'}$). In our supercell MD, only the $\Gamma$-point phonon mode is available, hence $\dt{\mathbf{R}} = \dt{U}_{\mathbf{R}}(t)$ (the dot indicates time derivative). Here, we make an approximation of $q$-point phonon mode based on the $\Gamma$-point phonon mode, as a result, we have:

\begin{align}
    \dt{\mathbf{R}} & =\frac{1}{\sqrt{N_{\mathbf{q}}}}\sum_{\mathbf{q}} \dt{U}_{\mathbf{R}}(t) e^{i\mathbf{q}\cdot\mathbf{r}} e^{i\theta_\mathbf{q}}
\end{align}

\noindent and $\theta_{\mathbf{q}}$ is a random phase. $N_{\mathbf{q}}$ is the number of $\mathbf{q}$ points. Plug this equation into equ. 1 with $\left|\phi_{i\mathbf{k}}(t)\right>=\left|u_{i\mathbf{k}}(t)\right>e^{i\mathbf{k}\cdot\mathbf{r}}$ and we can introduce cross $k$-point couplings. More specifically, using numerical derivative, assume $\Delta H=H(t+\Delta T)-H(t) = \frac{\partial H}{\partial \mathbf{R}} \dt{\mathbf{R}} \Delta T$, we can have:

\begin{align}
    V_{\mathbf{k}i,\mathbf{k'}j}(t) & = \left\langle \phi_{i\mathbf{k}}(t) \middle| \frac{\partial \phi_{j\mathbf{k'}}(t)}{\partial t} \right\rangle \nonumber\\
                                    & \approx \frac{1}{\Delta T\sqrt{N_{\mathbf{q}}}} \frac{1}{\epsilon_{j\mathbf{k'}}(t+\Delta T)-\epsilon_{i\mathbf{k}}(t)} \left\langle u_{i\mathbf{k}}(t)\middle|\Delta H\middle| u_{j\mathbf{k'}}(t+\Delta T)\right\rangle e^{i\theta_{\mathbf{k'-k}}}
\end{align}

\noindent where $\Delta T$ is the time step, $\epsilon_{i\mathbf{k}}$ is the eigen energy of state $\ket{\phi_{i\mathbf{k}}}$ of the supercell, $N_{\mathbf{q}}$ is the number of phonon wavevectors, which equals the number of $k$ points sampled in the supercell's Brillouin zone. Here, the momentum conservation is apparent during the derivation (see Supplementary Information (SI) for derivation details) and only the phonon modes wavevector satisfying $\mathbf{q}=\mathbf{k}-\mathbf{k'}$ are contributing. Note, the $\left\langle u_{i\mathbf{k}}(t)\middle|\Delta H\middle| u_{j\mathbf{k'}}(t+\Delta T)\right\rangle$ can be evaluated on the fly using $\left<u_{i\mathbf{k}}(t)\middle | u_{j\mathbf{k'}}(t+\Delta T)\right>$ and $\epsilon_{i\mathbf{k}}(t)$, $\epsilon_{j\mathbf{k'}}(t+\Delta T)$ etc. In actual caculations, we find that $\theta_{\mathbf{k'-k}}$ doesn't affect the the results significantly, thus we set it to zero. 

Following the CPA, in practical implementations, an \textit{ab initio} MD simulation of a fairly large supercell with a multi $k$-point sampling is performed (the supercell can't be too small since we want the trajectory to represent enough phonon modes). Here, the plane-wave density functional code PWmat~\cite{Jia13p9,Jia13p102} is used. During each MD step, when two eigen states $\ket{\phi_{i\mathbf{k}}(t)}$ and $\ket{\phi_{j\mathbf{k'}}(t+\Delta T)}$ have the same $k$ points (i.e. $\mathbf{k}=\mathbf{k'}$), $V_{ij}(t)=\left\langle \phi_i(t)\middle |\phi_j(t+\Delta T) \right\rangle$ is evaluated following the same approach presented in the previous works but subject to a $\frac{1}{\sqrt{N_{\mathbf{q}}}}$ prefactor~\cite{Kang19p224303,Zheng19p6174}; when $\mathbf{k}\neq \mathbf{k'}$, the above equ. 3 is evaluated. As shown in SI, these couplings are further modified to eliminate the possible spurious large values resulting from an accidental degeneracy between $\epsilon_{i\mathbf{k}}$ and $\epsilon_{j\mathbf{k'}}$ during the MD (see SI). The obtained non-adiabatic couplings $V_{i\mathbf{k},j\mathbf{k'}}$ are then applied to the DISH scheme (we can also use the equivalent P-matrix method (see SI Fig. 6)), where the wavefunction (or density matrix) of the excited carrier is expanded on the basis of adiabatic states including the states from all the $k$ points ($\ket{\psi(t)}=\sum_{i\mathbf{k}}c_{i\mathbf{k}}(t)\ket{\phi_{i\mathbf{k}}(t)}$). The wavefunction is evolved following the Schr\"{o}dinger's equation with the time-dependent Hamiltonians ($H(t)$) built from the non-adiabatic couplings. All we changed here is to provide the cross $k$-point couplings $V_{i\mathbf{k},j\mathbf{k'}}$. 

In order to verify the multi $k$-point NAMD scheme, shown in Fig. 1 is a comparison of the hot electron cooling NAMD simulations between a  2$\times$2$\times$2 Silicon cubic supercell (64 atoms) with 2$\times$2$\times$2 $k$-point sampling and a 4$\times$4$\times$4 cubic supercell (512 atoms) with the single $\Gamma$-point sampling. We apply the previous standard DISH without cross-$k$ transitions to the 4$\times$4$\times$4 supercell (since there is only one $k$ point) and apply the aforementioned multi-$k$ scheme to the 2$\times$2$\times$2 supercell. By setting an excited electron at the approximately same energies (around 0.925 eV) above the conduction band minimum (CBM) for these two systems, NAMD simulations are  performed to monitor the averaged energies ($\left<E(t)\right>=\sum_{i\mathbf{k}}\left|c_{i\mathbf{k}}(t)\right|^2\epsilon_{i\mathbf{k}}(t)$ where $c_{i\mathbf{k}}$ is the wavefunction expansion coefficients as $\ket{\psi(t)}=\sum_{i\mathbf{k}}c_{i\mathbf{k}}(t)\ket{\phi_{i\mathbf{k}}(t)}$) of the electron as a function of time. Here, each surface-hopping NAMD trajectory is an independent stochastic simulation and the final statistical mean energy is averaged over all the trajectories ($\Bar{E}(t)=1/N\sum_{i=1}^N \left<E_i(t)\right>$ with $N$ the number of trajectories). In this work, at least 100 independent trajectories are carried out to yield the averaged energy (see SI Fig.8 for convergence of number of trajectories). By turning on the cross-$k$ transitions in the 2$\times$2$\times$2 supercell, in principle, its electronic structure and the hot carrier cooling process should be the same to the 4$\times$4$\times$4 supercell case.  Shown in this figure, as we expect, these two simulations have almost the same energy-cooling profiles along the whole thermalization process. This result demonstrates the correctness of our method. In practice, the main computational cost of NAMD with the CPA is the \textit{ab initio} MD part. Compared with the large-supercell single $k$-point MD, the small supercell MD with multiple $k$-point reduces the computational cost significantly due to the $N^3$ scaling of the DFT calculations. Meanwhile, we also test the simulation by turning off the cross-$k$ transitions for the 2$\times$2$\times$2 supercell with 2$\times$2$\times$2 $k$-point grid (black line in Fig.1). The cooling is significantly impeded. In this case, the carrier can only populate the states within one $k$ point and large energy gaps are present in the energy spectrum of single $k$-point (see SI Fig.5).

\begin{figure}
\centering
\includegraphics[width=0.8\columnwidth]{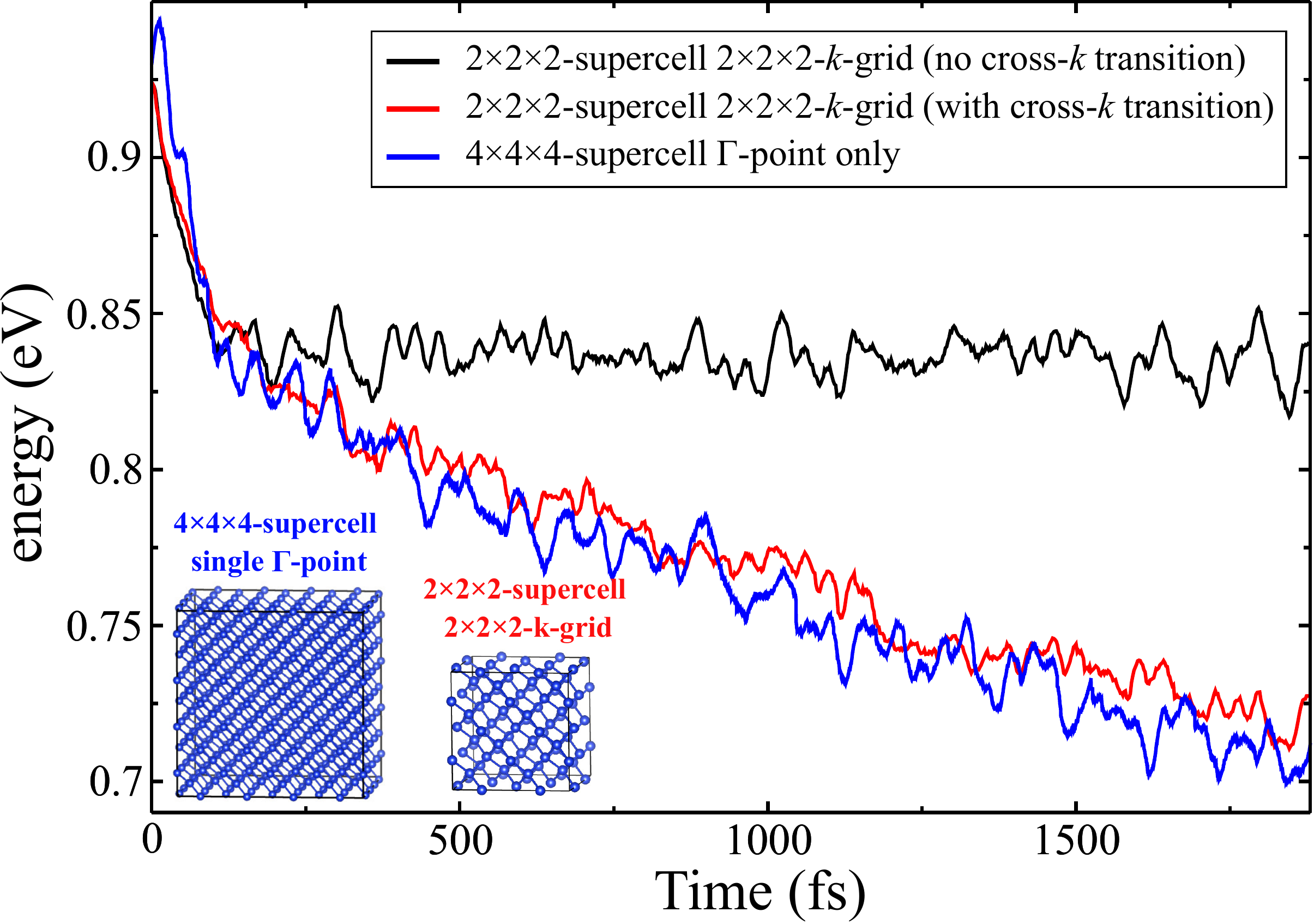}
\caption{A comparison of the averaged energy (averaged over trajectories) of an excited electron thermalization process simulated by NAMD between 4$\times$4$\times$4 Si cubic supercell (512 atoms) with single $\Gamma$-point, 2$\times$2$\times$2 supercell (64 atoms) with 2$\times$2$\times$2 $k$-point sampling enabling cross-$k$ transitions. A simulation for 2$\times$2$\times$2 supercell with 2$\times$2$\times$2 $k$-point sampling but \textbf{without} cross-$k$ transitions is also shown. To correct the finite-size effect to the eigen-energy fluctuation, the 2$\times$2$\times$2 supercell \textit{ab initio} MD is carried out at the temperature of 50K and the 4$\times$4$\times$4 supercell case is at 300K (see main text). Energy 0 of the $y$-axis is defined as the CBM of both systems. The starting energy for this electron is set around 0.925 eV for both systems.  }
\label{fig1}
\end{figure}

Here, we like to discuss the finite-size effect on the eigen energy fluctuations during the MD simulation. As shown in SI Fig.1, a 2$\times$2$\times$2 supercell with a 2$\times$2$\times$2 $k$-grid sampling and 4$\times$4$\times$4 supercell MD at the same temperature do not have the same eigen energy fluctuation: $\Delta \epsilon_i(t) \equiv \sqrt{\left<\left|\epsilon_i(t)-\epsilon_{\rm ave}\right|^2\right>}$ ($\epsilon_{\rm ave}$ is the average energy along the MD trajectory for this state). As shown in SI, $\Delta \epsilon \propto \sqrt{\frac{T}{N}}$ ($N$ is the supercell size and $T$ is the simulation temperature). Thus smaller the supercell, larger the eigen energy fluctuation. Such fluctuation is artificial. In order to have a correct comparison between the 2$\times$2$\times$2 supercell and the 4$\times$4$\times$4 supercell, we should adjust the MD thermostat of the 2$\times$2$\times$2 supercell to a lower temperature ($T=38$ K), but correct the non-adiabatic coupling constant of NAMD performed in this small supercell by a factor of $\sqrt{\frac{T_0}{T}}$ where $T_0=300$ K is the target temperature of the 4$\times$4$\times$4-supercell MD and NAMD simulations (SI Fig.2). However, further correction is possible. The above $\Delta\epsilon\propto \sqrt{\frac{T}{N}}$ scaling is under the assumption that the electron wavefunctions are completely delocalized. But as shown in SI Fig. 3, in the 4$\times$4$\times$4 supercell MD, the wavefunctions are already effectively localized in a smaller space than the whole space of the supercell. The averaged effective size of the supercell wavefunction is then calculated. Based on this effective size, we have used a final $T=50$ K in the 2$\times$2$\times$2 supercell MD simulation, and have used that in the final comparison between 2$\times$2$\times$2 supercell and 4$\times$4$\times$4 supercell as shown in Fig. 1.

\begin{figure}
\centering
\includegraphics[width=0.8\columnwidth]{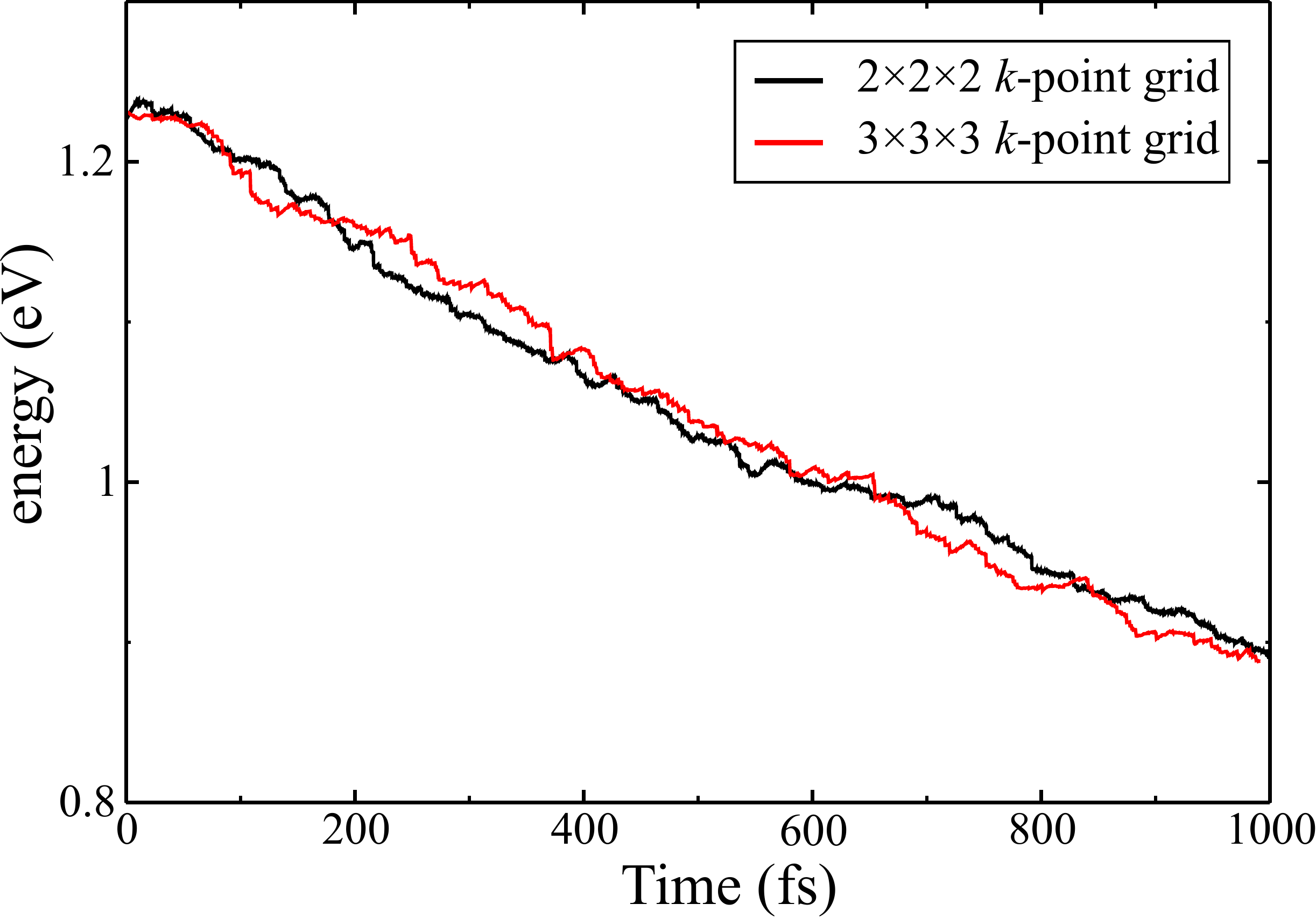}
\caption{NAMD simulation of the electron thermalizations with the 2$\times$2$\times$2 and 3$\times$3$\times$3 $k$-point grids carried on the 2$\times$2$\times$2 supercell. The electron is initially set around 1.2 eV above the CBM for these two types of $k$-point samplings. Energy 0 of $y$-axis is defined as the CBM of the system.}
\label{fig2}
\end{figure}

We next test the convergence of the $k$-point grids by using the same 2$\times$2$\times$2 supercell. Fig. 2 compares the two $k$-point sampling grids performed on the 2$\times$2$\times$2 silicon cubic supercell by setting the excited electron around 1.2 eV above the CBM. It is clear to see that even a 2$\times$2$\times$2 $k$-point sampling is already enough to simulate the electron thermalization process in such a system. A higher $k$-point grid will not change the picture significantly. Here, extra attentions should be applied to how the $k$ points are sampled. If following the Monkhorst-Pack sampling without a shift, high symmetric $k$ points with highly degenerate states are usually sampled, which also tends to create large artificial energy gaps. Therefore, in our MD simulation, a rigid random $k$-point shift is applied to the $k$-point grids to minimize the effect of state degeneracies (see SI Fig. 5).

So far what still missing is the role of acoustic phonon modes. When we multiply a factor of $e^{i\mathbf{q\cdot r}}$ to the $\Gamma$-point phonon modes to approximate the $\mathbf{q}$-point phonon modes, since the fixed box simulation does not contain the $\mathbf{q}= 0$ acoustic modes, our approximated phonon modes by multiplying $e^{i\mathbf{q}\cdot\mathbf{r}}$ also does not contain the corresponding acoustic modes. In Bardeen and Shockley's original paper about the deformation potential~\cite{Bardeen50p72}, the carrier mobility is computed purely based on the acoustic phonon modes induced crystal dilations. Here, on top of the aforementioned formalism, a deformation potential term $V^{\rm DP}_{i\mathbf{k},j\mathbf{k'}}$ is further added to the non-adiabatic couplings (equ. 4). Equ. 5 is the expression of the coupling strength contributed by the deformation potential~\cite{Bardeen50p72,Murphy-Armando10p869}, which is then added to $V_{\mathbf{k}i,\mathbf{k'}j}(t)$ (only when $\mathbf{k}\neq\mathbf{k'}$) to yield the total couplings $V^{\rm tot}_{\mathbf{k}i,\mathbf{k'}j}(t)$ used in NAMD:

\begin{align}
    V^{\rm tot}_{i\mathbf{k},j\mathbf{k'}}(t) & = V_{i\mathbf{k},j\mathbf{k'}}(t)  + V^{\rm DP}_{i\mathbf{k},j\mathbf{k'}}(t), \;\text{ when $\mathbf{k}\neq\mathbf{k'}$} \\
    V^{\rm DP}_{i\mathbf{k},j\mathbf{k'}}(t) & = \frac{1}{\Delta T\sqrt{N_{\mathbf{q}}}} \frac{1}{\epsilon_{j\mathbf{k'}}(t+\Delta T)-\epsilon_{i\mathbf{k}}(t)} M^{\rm DP}_{i\mathbf{k},j\mathbf{k'}} \\
    M^{\rm DP}_{i\mathbf{k},j\mathbf{k'}}(t) & = E_1 \sqrt{\frac{k_bT}{2M\omega_{\mathbf{q}}^2}}\left|\mathbf{q}\right| \omega_{\mathbf{q}} \Delta T \left<u_{i\mathbf{k}}(t)\middle|u_{j\mathbf{k'}}(t+\Delta T)\right>
\end{align}

\noindent where $E_1$ is the deformation potential factor and $\mathbf{q}=\mathbf{k}-\mathbf{k'}$. $M$ is the mass of silicon atom and $\omega_{\mathbf{q}}$ is the acoustic phonon mode energy which is cancelled. Here, we assume the acoustic mode is simply $e^{i\mathbf{q\cdot r}}$. We have followed Bardeen's original derivation to arrive in the matrix element in equ. 6. $E_1$ of silicon conduction bands is taken from Ref.~\citenum{Bardeen50p72}. Note, $E_1$ can be calculated \textit{ab initio-ly} by calculating the band gap deformation potential, then assigning $2/3$ to the conduction band and $1/3$ to the valence band~\cite{Li06p042104}. Although more accurate $E_1$ calculation might be possible, we find the final results does not depend on $E_1$ significantly, so a rough approximation is enough. For silicon, we have used $E_1=6.5$ eV. $\left<u_{i\mathbf{k}}(t)\middle|u_{j\mathbf{k'}}(t+\Delta T)\right>$ is computed to identify which $j\mathbf{k'}$ state belongs to the same band of state $i\mathbf{k}$ owing to the intra-band nature for the acoustic mode induced transitions. This term can be calculated on the fly during MD.

\begin{figure}
\centering
\includegraphics[width=0.8\columnwidth]{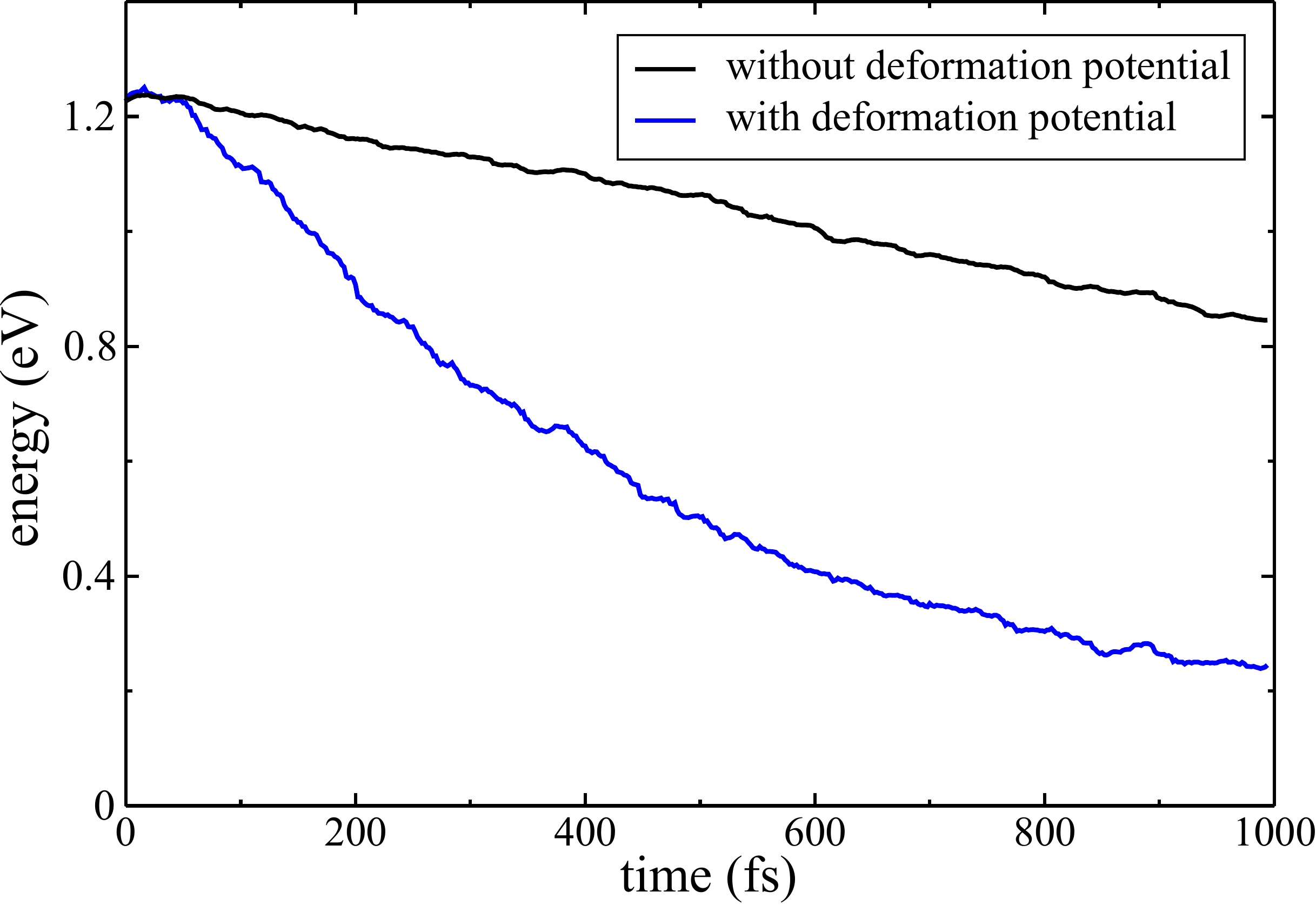}
\caption{NAMD simulation of electron thermalizations with the deformation potential in addition to cross-$k$ transitions. This calculation is performed on a 2$\times$2$\times$2 silicon supercell with 2$\times$2$\times$2 $k$-point grid. The initial position of the excited electron is set around 1.2 eV above the CBM.}
\label{fig3}
\end{figure}

In our actual NAMD simulations, the DISH algorithm implementation is based on the overlap matrix element $S_{i\mathbf{k},j\mathbf{k'}}=\left<u_{i\mathbf{k}}(t)\middle|u_{j\mathbf{k'}}(t+\Delta T)\right>$ (here we should consider $\mathbf{k}$ as a state index in a large supercell, so there are overlaps between state $i\mathbf{k}$ and $j\mathbf{k'}$). The above procedures give us the approximated $V^{\rm tot}_{i\mathbf{k},j\mathbf{k'}}$, which is equivalent to the overlap matrix element $S$ subject to a $\Delta T$ factor. However, the overlap matrix $S_{i\mathbf{k},j\mathbf{k'}}$ should satisfy some normalization rules which might be missing due to the approximation. We have applied an orthonormalization procedure as described in SI section 7 before using it in DISH algorithm to carry out the final simulations. Shown in Fig. 3 is the comparison of electron thermalization with and without the deformation potential. By adding the deformation potential, the electron cooling rate is accelerated significantly. Along the whole thermalization process, the energy relaxation rate is not fixed: for the initial 200 fs, the cooling rate can reach around 2 eV/ps, but for the last 400 fs (from 600 fs to 1000 fs), the cooling rate reduces to around 0.5 eV/ps. The relative change of cooling rate is directly related to the density of states as shown in SI Fig. 5. Here, we want to comment that the added deformation potential is not to bring down the energy of excited electrons directly. Instead, it enhances the momentum equilibration between different $k$ points (but with similar energies) to multiply the cooling channels. As shown in SI Fig. 11, by adding the deformation potential, the population of the excited carrier wavefunction spreads to more states. Thus, while the optical phonon modes might thermalize the electron by reducing its energy, the supercell acoustic phonon modes prompt the distribution of the carriers among states with similar energies. In this case, the energy-cooling rate is actually not sensitive to the exact value of deformation potential factor $E_1$ (SI Fig.10). Meanwhile, we also test the convergence of $k$-point grids after adding the deformation potential. As shown in SI Fig. 9, the 2$\times$2$\times$2 $k$-point grid agrees well with the 3$\times$3$\times$3 $k$-point grid when the deformation potential is added.

\begin{figure}
\centering
\includegraphics[width=0.8\columnwidth]{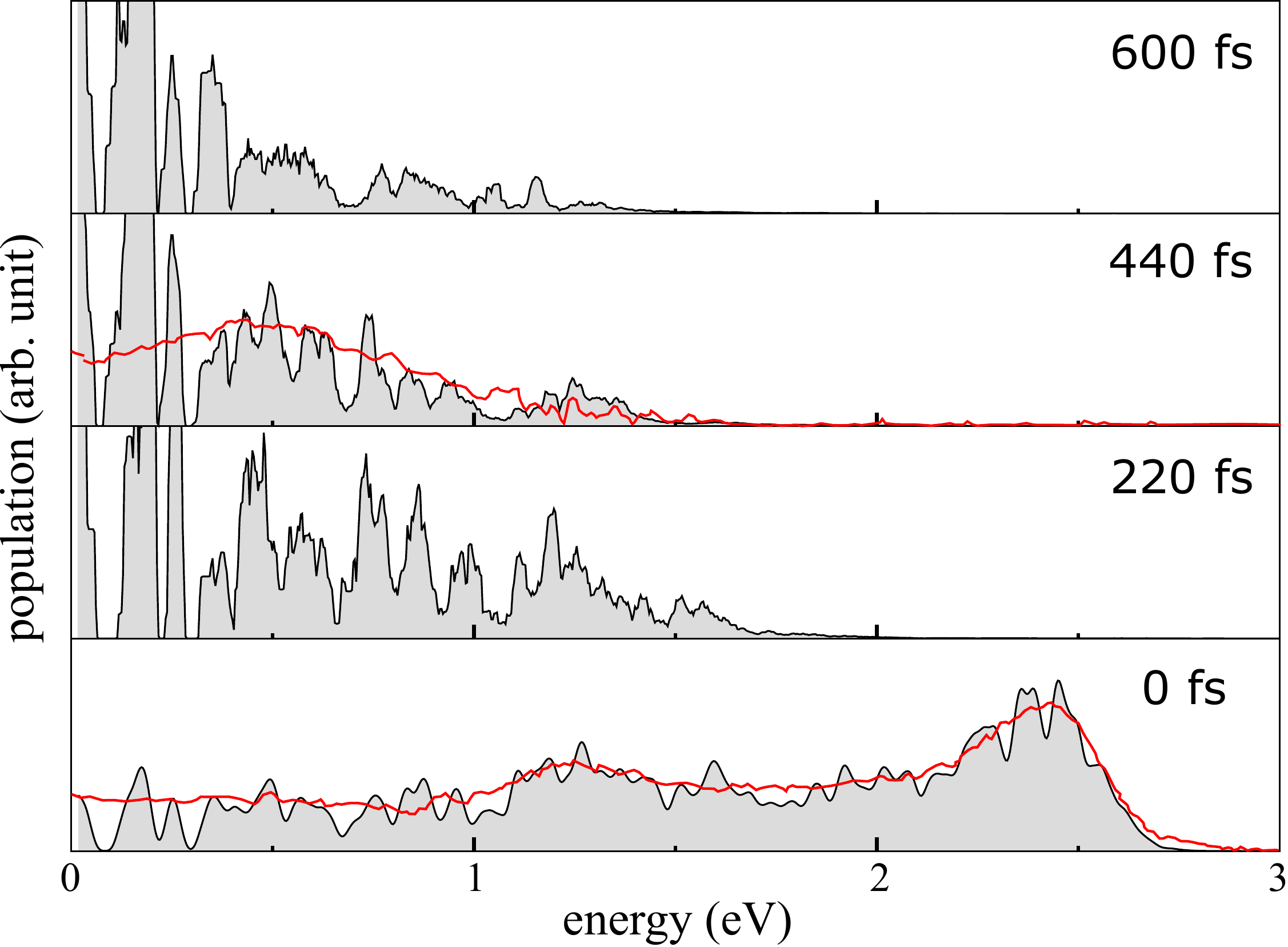}
\caption{A NAMD simulated populations of hot electrons in bulk silicon. Time $t=0$ fs is set manually to be consistent to the measured electron population from Ref.~\citenum{Tanimura19p035201} Fig. 4. The red lines are reproduced from Ref.~\citenum{Tanimura19p035201}. Energy 0 of $x$-axis is defined as the CBM. }
\label{fig4}
\end{figure}

A well calibrated carrier relaxation experiment is very challenging. Only recently such experiments become available. Recently, Tanimura et al has carried such an experiment for bulk silicon hot carrier thermalization~\cite{Tanimura19p035201}. After the excitation of a short pulse pump light, a transient detection of electron populations was obtained by the probing light. In our calculation, we manually populate the states based on the population measured experimentally~\cite{Tanimura19p035201} shortly after the pump (around 10 fs). With this initial population, we carry out the NAMD and compare the calculated population change after a finite time with the measurements (the population is defined as the occupation of wavefunction to the eigen states: $\left|c_{i\mathbf{k}}(t)\right|^2$ with $\ket{\psi(t)}=\sum_{i\mathbf{k}}c_{i\mathbf{k}}(t)\ket{\phi_{i\mathbf{k}}(t)}$). Shown in Fig. 4 is the time evolution of the state population for time $t=0$ fs, $t=220$ fs, $t=440$ fs, and $t=600$ fs. As shown in the figure, at $t=440$ fs when the experimental result is available, the agreement is very good~\cite{Tanimura19p035201}. We do note that near the CBM (within 0.2 eV), we have slightly larger population. This is because experimentally there are defect states and surface states below the CBM, which attract the electron population~\cite{Tanimura19p035201}. 

In summary, we propose a first-principle method to directly simulate the hot carrier thermalization and transfer process in periodic bulk systems. An approximated cross $k$-point coupling terms is introduced based on the conventional MD. Using the bulk silicon as an example, by sampling a multi $k$-point grids during the \textit{ab initio} MD and computing the cross-$k$ transition non-adiabatic couplings, a NAMD simulation performed on a small supercell with a multi $k$-point grid agrees well with the regular NAMD performed on a large supercell with a single $\Gamma$ point. By adding the deformation potential, our simulated bulk silicon hot electron thermalization process agrees excellently with the most recent experiments. This method can be applied not only to the hot carrier cooling but also to various carrier transfer phenomena in solid states as well as 2D materials, which provides an efficient tool to investigate the ultrafast dynamic processes. 

{\large \textbf{Acknowledgment}} 

F. Zheng was supported by the ShanghaiTech University starting package. L. W. Wang was supported by the key research program of the Chinese Academy of Sciences, Grant No. ZDBS-SSW-WHCOO2. The computational support was provided by the high performance computing facility in ShanghaiTech University.  

%

\end{document}



\title{{\Large Supplementary Information:} A multi $k$-point nonadiabatic molecular dynamics for periodic systems}

\author{Fan Zheng}
 \affiliation{School of physical science and technology, ShanghaiTech University, Shanghai 201210, China. }
 \email{zhengfan@shanghaitech.edu.cn}
\author{Lin-wang Wang}
 \affiliation{State key laboratory of superlattices and microstructures, Institute of semiconductors, Chinese Academy of Science, Beijing 100083, China.}
 \email{lwwang@semi.ac.cn}

\maketitle

\begin{itemize}
    \item {\large \textbf{Derivation of cross-$k$ non-adiabatic couplings}}
    \item {\large \textbf{\textit{ab initio} MD calculation details}}
    \item {\large \textbf{Finite-size effect for eigen energy fluctuation during MD}}
    \item {\large \textbf{Eigen states from $\Gamma$ point only and DOS}}
    \item {\large \textbf{Decoherence induced surface hopping (DISH)}}
    \item {\large \textbf{Decoherence time}}
    \item {\large \textbf{Orthonormalization of overlap matrix}}
    \item {\large \textbf{Convergence of number of trajectories}}
    \item {\large \textbf{$k$-point grid convergence with DP}}
    \item {\large \textbf{Different deformation potential factor $E_1$}}
    \item {\large \textbf{An example illustration of $k$-resolved population}}
\end{itemize}

\newpage

\section{1. Derivation of cross-$k$ non-adiabatic couplings}

From the equ. 1 in the main text, we can derive the cross-$k$ ($\mathbf{k}\neq\mathbf{k'}$) non-adiabatic coupling as:

\begin{align}
    V_{i\mathbf{k},j\mathbf{k'}}(t) & = \left\langle \phi_{i\mathbf{k}}(t) \middle| \frac{\partial \phi_{j\mathbf{k'}}(t)}{\partial t} \right\rangle \\
                                    & = \sum_{\mathbf{R}} \frac{1}{\epsilon_{j\mathbf{k'}}-\epsilon_{i\mathbf{k}}} \left\langle\phi_{i\mathbf{k}}(t)\middle|\frac{\partial H}{\partial \mathbf{R}}\cdot \dt{\mathbf{R}}\middle|\phi_{j\mathbf{k'}}(t)\right\rangle\\
                                    & = \frac{1}{\sqrt{N_{\mathbf{q}}}} \sum_{\mathbf{R}} \frac{1}{\epsilon_{j\mathbf{k'}}-\epsilon_{i\mathbf{k}}} \left\langle u_{i\mathbf{k}}(t)\middle|\frac{\partial H}{\partial \mathbf{R}}e^{i(\mathbf{k'-k+q})\cdot\mathbf{r}}\middle| u_{j\mathbf{k'}}(t)\right\rangle \dt{U}_{\mathbf{q},\mathbf{R}}(t)  e^{i\theta_{\mathbf{k'-k}}}  \\
                                    & = \frac{1}{\sqrt{N_{\mathbf{q}}}} \sum_{\mathbf{R}} \frac{1}{\epsilon_{j\mathbf{k'}}-\epsilon_{i\mathbf{k}}} \left\langle u_{i\mathbf{k}}(t)\middle|\frac{\partial H}{\partial \mathbf{R}}\middle| u_{j\mathbf{k'}}(t)\right\rangle \dt{U}_{\mathbf{k'-k},\mathbf{R}}(t)  e^{i\theta_{\mathbf{k'-k}}}  \\
                                    & \approx \frac{1}{\Delta T\sqrt{N_{\mathbf{q}}}} \frac{1}{\epsilon_{j\mathbf{k'}}(t+\Delta T)-\epsilon_{i\mathbf{k}}(t)} \left\langle u_{i\mathbf{k}}(t)\middle|\Delta H\middle| u_{j\mathbf{k'}}(t+\Delta T)\right\rangle  e^{i\theta_{\mathbf{k'-k}}}
\end{align}

\noindent where $\mathbf{R}$ is the atomic position in the supercell and $\mathbf{q}$ is the phonon wavevector. Equ. 1 to equ. 2 is based on the expansion of $\frac{\partial}{\partial \mathbf{R}}\left<\phi_{i}(t)\middle|H(t)\middle|\phi_{j}(t)\right>=0$ ($i\neq j$). Here, we want to emphasize that the inner product between the adiabatic states of different wavevectors in equ. 1 will not vanish even when $\mathbf{k}\neq\mathbf{k'}$. Instead, it contains the information of electron-phonon interactions. The time evolution of eigen function $\ket{\phi_{j\mathbf{k'}}}$ changes not only the periodic part $u_{j\mathbf{k'}}$ ($\left|\phi_{i\mathbf{k}}(t)\right>=\left|u_{i\mathbf{k}}(t)\right>e^{i\mathbf{k}\cdot\mathbf{r}}$), but also the wavevector $\mathbf{k'}$, since the phonon modes will re-modulate the periodicity of electron wavefunctions. Equ. 2 to 3 is based on the approximated expansion of an atom's trajectory to the phonon modes: $\dt{\mathbf{R}}=\frac{1}{\sqrt{N_{\mathbf{q}}}}\sum_{\mathbf{q}} \dt{U}_{\mathbf{q},\mathbf{R}}(t) e^{i\mathbf{q}\cdot\mathbf{r}} e^{i\theta_{\mathbf{q}}}$ (phonon mode index is not shown for simplicity). It is apparent that given the term $e^{i\mathbf{q}\cdot\mathbf{r}}$, only when $\mathbf{q}=\mathbf{k}-\mathbf{k'}$, the integral in equ. 3 will not vanish, which also satisfies the momentum conservation. From equ. 4 to 5, we assume that the phonon-mode vectors of different phonon wavevectors are similar, which can be replaced by the MD atomic trajectories $\Delta \mathbf{R}$ i.e. $\sum_{\mathbf{R}} \frac{\partial H}{\partial \mathbf{R}} \dt{U}_{\mathbf{R}}\Delta T = \Delta H$ ($\Delta H\equiv H(t+\Delta T) - H(t)$). Equ. 5 is written with a discrete time in order to evaluate this term on the fly during a MD simulation. Such discrete time expression can also be derived as ($\mathbf{k}\neq\mathbf{k'}$):

\begin{align}
    V_{i\mathbf{k},j\mathbf{k'}}(t) & = \left\langle \phi_{i\mathbf{k}}(t) \middle| \frac{\partial \phi_{j\mathbf{k'}}(t)}{\partial t} \right\rangle \nonumber\\
                                    & \approx \sum_{\mathbf{R}} \left<\phi_{i\mathbf{k}}(t)\middle|\frac{\phi_{j\mathbf{k'}}(t+\Delta T)}{\Delta \mathbf{R}}\dt{\mathbf{R}}\right>  \nonumber\\
                                    & = \sum_{\mathbf{R}} \frac{\epsilon_{j\mathbf{k'}}(t+\Delta T)\left<\phi_{i\mathbf{k}}(t)\middle|\phi_{j\mathbf{k'}}(t+\Delta T)\dt{\mathbf{R}}\right>-\epsilon_{i\mathbf{k}}(t)\left<\phi_{i\mathbf{k}}(t)\middle|\phi_{j\mathbf{k'}}(t+\Delta T)\dt{\mathbf{R}}\right>}{\left(\epsilon_{j\mathbf{k'}}(t+\Delta T)-\epsilon_{i\mathbf{k}}(t)\right)\Delta \mathbf{R}} \nonumber\\
                                    & = \sum_{\mathbf{R}} \frac{\left<\phi_{i\mathbf{k}}(t)\middle|\frac{H(t+\Delta T)}{\Delta\mathbf{R}}\dt{\mathbf{R}}\middle|\phi_{j\mathbf{k'}}(t+\Delta T)\right>-\left<\phi_{i\mathbf{k}}(t)\middle|\frac{H(t)}{\Delta\mathbf{R}}\dt{\mathbf{R}}\middle|\phi_{j\mathbf{k'}}(t+\Delta T)\right>}{\epsilon_{j\mathbf{k'}}(t+\Delta T)-\epsilon_{i\mathbf{k}}(t)} \nonumber\\
                                    & = \sum_{\mathbf{R}} \frac{1}{\epsilon_{j\mathbf{k'}}(t+\Delta T)-\epsilon_{i\mathbf{k}}(t)} \left<\phi_{i\mathbf{k}}(t)\middle|\frac{\Delta H}{\Delta\mathbf{R}}\dt{\mathbf{R}}\middle|\phi_{j\mathbf{k'}}(t+\Delta T)\right> 
\end{align}

\noindent where $\Delta H \equiv H(t+\Delta T)-H(t)$. It arrives at the above equ. 2. This is also how the cross-$k$ matrix element is computed during MD.

\section{2. \textit{ab initio} MD calculation details}

All the first-principle calculations including structural relaxation, self-consistent field calculation, and \textit{ab initio} MD are performed with the plane-wave pseudopotential DFT package PWmat~\cite{Jia13p9,Jia13p102} with the generalized gradient approximation (GGA) functional~\cite{Perdew96p3865}. The silicon SG15 pseudopotential~\cite{Hamann13p085117} with 45 Ryd plane-wave energy cut-off are used to converge the charge density of the bulk silicon. The \textit{ab initio} MD simulations of different supercell sizes at different temperatures are performed with time step 2 fs. During MD, the quantities such as $\left<u_{i\mathbf{k}}(t)\middle|u_{j\mathbf{k}}(t+\Delta T)\right>$, $\left<u_{i\mathbf{k}}(t)\middle|\Delta H(t)\middle|u_{j\mathbf{k'}}(t+\Delta T)\right>$, and $\epsilon_{i\mathbf{k}}(t)$ are saved to files for the post-processing NAMD. The starting points of trajectories used in NAMD is chosen when the system temperatures are reaching equilibrium for the target temperature.

\section{3. Finite-size effect for eigen energy fluctuation during MD}

When comparing the hot carrier thermalization process between the 2$\times$2$\times$2 silicon cubic supercell (64 atoms) with a 2$\times$2$\times$2 $k$-point grid and the 4$\times$4$\times$4 supercell (512 atoms) with a single $\Gamma$ $k$-point, it is necessary to ensure that the electronic structure of these two systems are similar for easy comparisons. Shown in SI Fig. 1 is the eigen energies evolution of all the $k$ points and states during the MD. Apparently, the two systems at the same temperature yield quite different eigen energy evolutions, particularly for the oscillation magnitude of the eigen energies. This can be explained by the following derivation. 

\begin{figure}
\centering
\includegraphics[width=1\columnwidth]{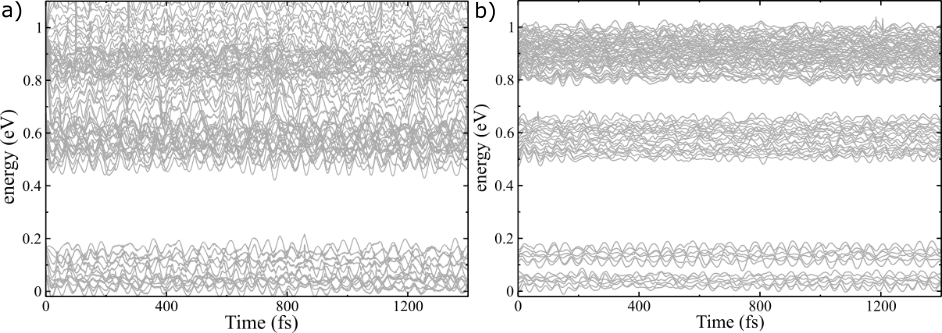}
\caption{Plot of eigen energies from \textit{ab initio} MD trajectories of all the $k$ points and states from a) 2$\times$2$\times$2 silicon supercell (2$\times$2$\times$2 $k$-point grid) at 300K and b) 4$\times$4$\times$4 silicon supercell (single $\Gamma$ point) at 300K. Zero of $y$-axis is defined as the CBM of the system.}
\label{sfig1}
\end{figure}

Assuming each unit cell only contains one atom and we have $N$ such unit cells forming a supercell. The eigen energy evolution can be expressed as:

\begin{align}
    \Delta E_{i\mathbf{k}}(t) & = \left<\phi_{i\mathbf{k}}(t)\middle|\Delta H(t)\middle|\phi_{i\mathbf{k}}(t)\right>  \\
                        \Delta H(t) & = \sum_{\mathbf{R}} \frac{\partial V(t)}{\partial \mathbf{R}} \Delta \mathbf{R} \\
                  \Delta \mathbf{R} & = \sum_{\lambda,\mathbf{q}} \frac{1}{\sqrt{NM}} \Delta U_{\lambda\mathbf{q}} e^{i\omega_{\lambda\mathbf{q}}t} e^{i\mathbf{q}\cdot\mathbf{r}}
\end{align}

\noindent where summation in equ. 8 is summed for all atoms in the supercell, $M$ is the mass of the atom, $\lambda$ and $\mathbf{q}$ are the phonon mode and the wavevector of the supercell, $U_{\lambda\mathbf{q}}$ is the phonon mode vector, and $\omega_{\lambda\mathbf{q}}$ is the phonon energy. Meanwhile, the Bloch state can be expressed as $\ket{\phi_{i\mathbf{k}}}\equiv\frac{1}{\sqrt{N}}\ket{u_{i\mathbf{k}}}e^{i\mathbf{k}\cdot\mathbf{r}}$ with $\int_{\rm unit cell} u^\ast_{i\mathbf{k}}(\mathbf{r})u_{i\mathbf{k}}(\mathbf{r}) d\mathbf{r}=1$. If we plugin equ. 8 and 9 into equ. 7, we obtain:

\begin{align}
    \Delta E_{i\mathbf{k}}(t) & = \sum_{\mathbf{R}} \sum_{\lambda,\mathbf{q}} \left<\phi_{i\mathbf{k}}(t)\middle| \frac{\partial V(t)}{\partial \mathbf{R}} \Delta U_{\lambda\mathbf{q}}  e^{i\omega_{\lambda\mathbf{q}}t} e^{i\mathbf{q}\cdot\mathbf{r}} \middle|\phi_{i\mathbf{k}}(t)\right>  \\
                                    & =  \frac{1}{\sqrt{NM}} \frac{1}{N}\sum_{\mathbf{R}} \sum_{\lambda} \left<u_{i\mathbf{k}}(t)\middle| \frac{\partial V(t)}{\partial \mathbf{R}}   \middle|u_{i\mathbf{k}}(t)\right> \Delta U_{\lambda,\mathbf{q}=\Gamma}  e^{i\omega_{\lambda\mathbf{q}}t} 
\end{align}

Here, equ. 10 to equ. 11 needs the momentum conservation with only $\mathbf{q}=\Gamma$ phonon mode contributing. By assuming that the Bloch wavefunction periodic part $\ket{u_{i\mathbf{k}}}$ is fully delocalized in all $N$ unit cells and are generally the same, and $U_{\lambda,\mathbf{q}=\Gamma}$ is also similar for all unit cells. The summation of $\mathbf{R}$ in equ. 11 can be replaced by $N$ times and we obtain:

\begin{align}
    \Delta E_{i\mathbf{k}}(t) & = \frac{1}{\sqrt{NM}} \sum_{\lambda} \left<u_{i\mathbf{k}}(t)\middle| \frac{\partial V(t)}{\partial \mathbf{R}}   \middle|u_{i\mathbf{k}}(t)\right> \Delta  U_{\lambda,\mathbf{q}=\Gamma} e^{i\omega_{\lambda\mathbf{q}}t} \\
    \left<\left|\Delta E_{i\mathbf{k}}(t)\right|^2\right> & = \frac{1}{NM}\sum_{\lambda} \left|\left<u_{i\mathbf{k}}(t)\middle| \frac{\partial V(t)}{\partial \mathbf{R}}   \middle|u_{i\mathbf{k}}(t)\right>\right|^2 \left|\Delta U_{\lambda,\mathbf{q}=\Gamma}\right|^2
\end{align}

Meanwhile, the magnitude of phonon mode vector is related to temperature approximately as:

\begin{align}
    \frac{1}{2} k_b T & = \frac{1}{2} M \omega^2_{\lambda\mathbf{q}} U^2_{\lambda,\mathbf{q}} \\
                    \left|U\right|^2 & \propto \frac{k_bT}{M\omega^2_{\lambda\mathbf{q}}}
\end{align}

Therefore, the magnitude of $\sqrt{\left<\left|\Delta E_{i\mathbf{k}}(t)\right|^2\right>}$ is actually proportional to $\sqrt{\frac{T}{N}}$. When decreasing the size of supercell from 4$\times$4$\times$4 (64 cubic unit cells) to 2$\times$2$\times$2 (8 cubic unit cells), the MD temperature should be changed from 300K to 38K ($38{\rm K}=300{\rm K}/8$) to ensure a consistent energy oscillation magnitude. The above derivation will not be different if there are multiple atoms in a unit cell; an additional summation over such basis should be applied. Shown in SI Fig. 2 is the comparison between 2$\times$2$\times$2 and 4$\times$4$\times$4 supercells with MD temperature set as 38K and 300K, respectively. However, the electronic structures of these two systems still do not match. 2$\times$2$\times$2 supercell tends to show a smaller eigen energy oscillations than that of the 4$\times$4$\times$4 supercell, which suggests a higher MD temperature for the 2$\times$2$\times$2 supercell is needed. 

\begin{figure}
\centering
\includegraphics[width=1\columnwidth]{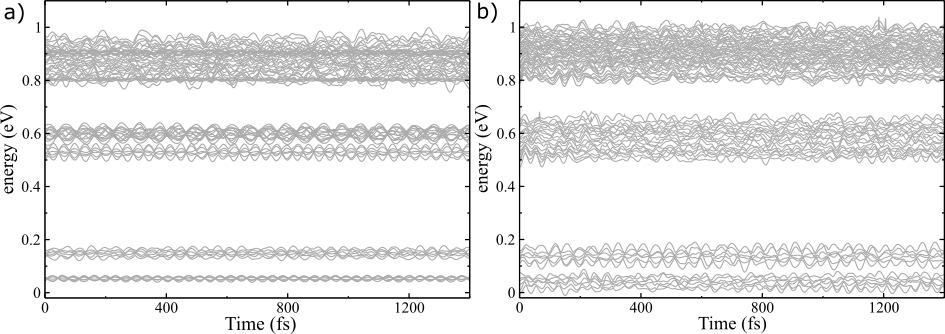}
\caption{Plot of eigen energies from MD trajectories of all $k$ points and all states from a) 2$\times$2$\times$2 silicon supercell at 38K and b) 4$\times$4$\times$4 silicon supercell at 300K. Zero of $y$-axis is defined as the CBM of the system.}
\label{sfig2}
\end{figure}

By reviewing the derivation above, we assume a fully delocalized wavefunction over all the unit cells in a supercell. But during an MD simulation, local strains tend to localize the wavefunction and the eigen states of the supercell may not necessarily spread to all the unit cells uniformly. SI Fig. 3 a) illustrates several eigen states from an MD trajectory of the 4$\times$4$\times$4 supercell. It is apparent that these eigen state wavefunctions are not circulating all the silicon atoms of the supercell. Instead, they are only occupying a localized spacial region with an effective number of unit cells $N'$ which is smaller than 64. If the temperature for the 2$\times$2$\times$2 supercell MD is computed based on the effective size $N'$, a better agreement of eigen energy evolutions between these two systems could be reached. Here we propose a way to quantify the effective size based on the charge density partition ratio:

\begin{figure}
\centering
\includegraphics[width=1\columnwidth]{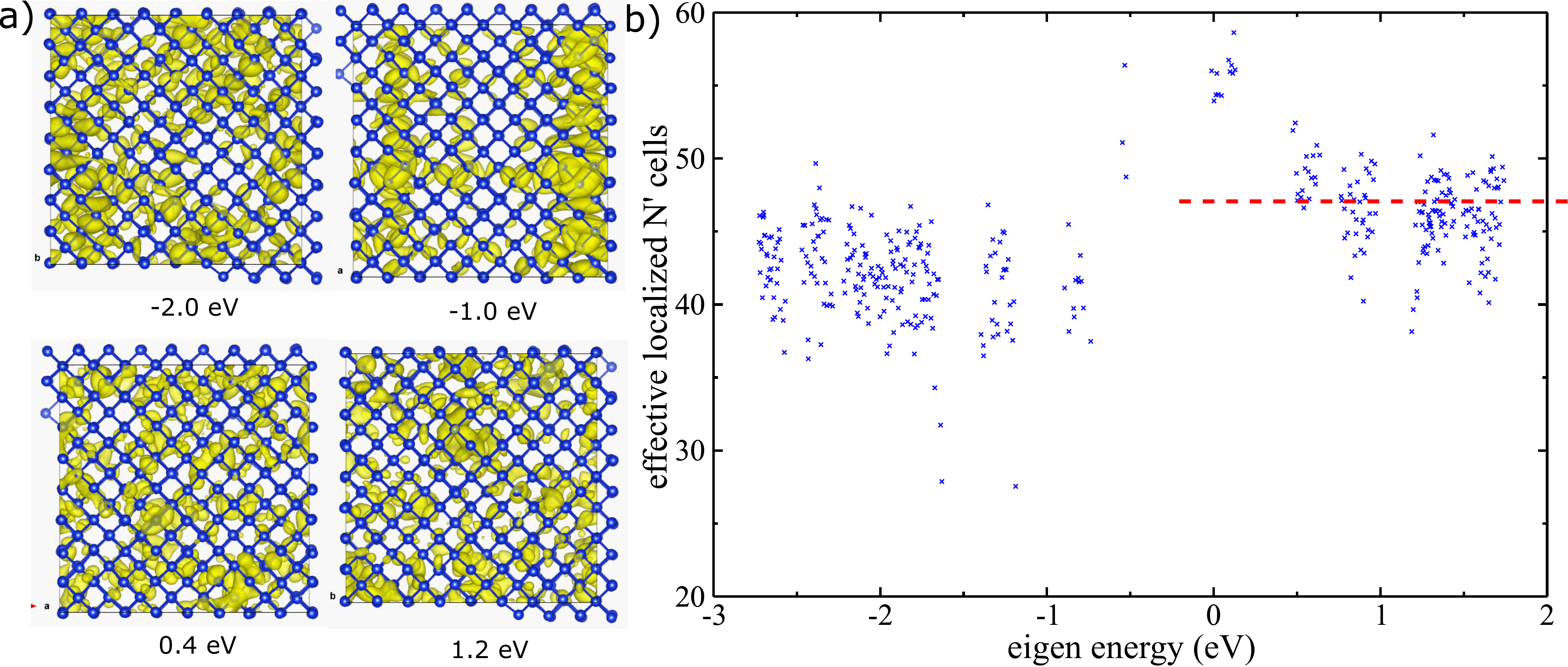}
\caption{a) Charge density of four states selected from a snapshot of a 300K 4$\times$4$\times$4 supercell MD simulation. The energy below each charge density plot indicates their eigen energies and 0 is the CBM state of silicon. b) The calculated effective size $N'$ for the 4$\times$4$\times$4 supercell. The 0 in $x$-axis is the CBM of silicon. }
\label{sfig3}
\end{figure}

\begin{align}
    \frac{1}{N'} = \frac{\int_V \phi_i^4 d\mathbf{r}}{\int_\Omega \phi_i^{'4} d\mathbf{r}}
\end{align}

\noindent where $V$ is integrated over the whole volume of supercells and $\Omega$ is for unit cells (8 atom cubic silicon unit cell), $\phi_i$ is the eigen state $i$ selected from an MD snapshot, and $\phi'_i$ is the wavefunction of this state but on a perfect equilibrium unit cell of silicon. 

Shown in SI Fig. 3 b) is the computed $N'$ for an MD snapshot of the 4$\times$4$\times$4 supercell performed at 300K. Since we are only interested in the conduction bands, here we use the averaged value of $N'$ which is around 48. In this case, the 2$\times$2$\times$2 supercell MD temperature is re-evaluated as 50 K. SI Fig. 4 shows the eigen energy evolution comparison between these two types of systems. By using the new MD temperature, the agreement of the eigen energy evolutions between the 2$\times$2$\times$2 and the 4$\times$4$\times$4 supercells improves significantly. With the similar electronic structures, it allows us to compare the NAMD simulations of these two systems. 

\begin{figure}
\centering
\includegraphics[width=1\columnwidth]{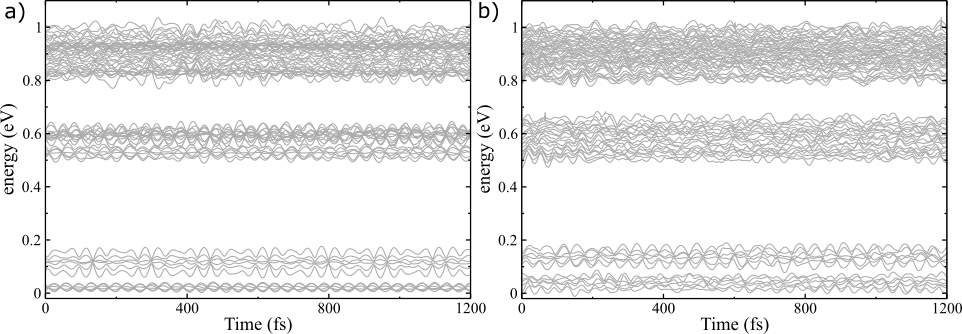}
\caption{Plot of eigen energies from MD trajectories of all $k$ points and all states from a) 2$\times$2$\times$2 silicon supercell at 50K and b) 4$\times$4$\times$4 silicon supercell  at 300K. Zero of $y$-axis is defined as the CBM of the system.}
\label{sfig4}
\end{figure}

\section{4. Eigen states from $\Gamma$ point only and from random shifted $k$ grid}

Shown in SI Fig. 5 is the eigen states evolution computed from the 2$\times$2$\times$2 supercell at $T=50$ K. By sampling a 2$\times$2$\times$2 $k$-point grid centered on $\Gamma$, SI Fig. 5 a) indicates the states from all the $k$ points and b) only the extracted $\Gamma$-point states are shown. c) shows the eigen states evolution computed from a 2$\times$2$\times$2 random shifted $k$-point grid. It is apparent that by sampling only $\Gamma$ point or high symmetric multi $k$-point, artificial large energy gaps are present which may hinder the excited carrier transfer or carrier thermalization. A random shifted $k$-grid will iron out these energy gaps effectively. 

\begin{figure}
\centering
\includegraphics[width=1\columnwidth]{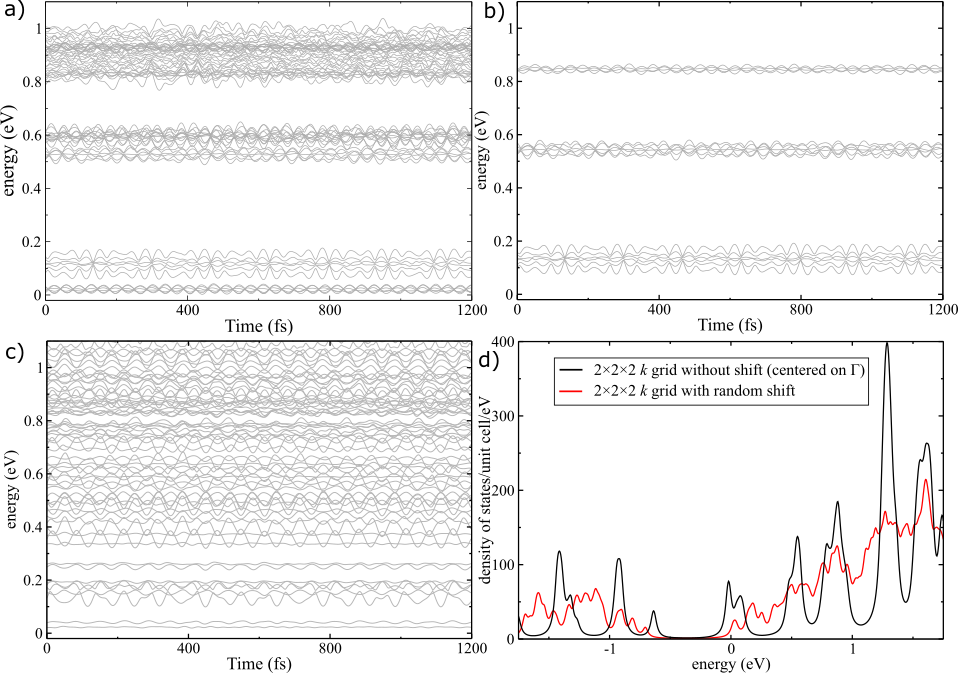}
\caption{Plot of eigen energies' evolution from the 2$\times$2$\times$2-supercell MD of a) states from all $k$ points (with 2$\times$2$\times$2 non-shifted $k$-point sampling), b) eigen states extracted from $\Gamma$ point only out of the 2$\times$2$\times$2 $k$-point sampling, c) states from a random shifted 2$\times$2$\times$2 $k$-grid, and d) a comparison of density of states (based on structure at time $t=0$) for shifted and non-shifted 2$\times$2$\times$2 $k$-point samplings.}
\label{sfig5}
\end{figure}

\section{5. Decoherence induced surface hopping (DISH)}

In this work, we implement the DISH method originally proposed in Ref.~\citenum{Jaeger12p22A545} for our NAMD simulation. The wavefunction of the excited carrier $\ket{\psi(t)}$ follows the Schr\"{o}dinger's equation and it is expanded with the adiabatic states (eigen states) $\ket{\phi_i(t)}$. The equation of motion for the coefficients used in the expansion is obtained as the following formula (index of $\mathbf{k}$ is not shown). The classical path approximation (CPA) is used, which assumes the ions' motion is always on the ground state potential energy surface. This is a fair approximation for large systems with heavy atoms and the wavefunctions are delocalized, particularly when we are only interested in the carrier dynamics. By using CPA, the NAMD simulation becomes a post-processing of \textit{ab initio} MD. The overlap between the two eigen states in equ. 19 is computed during the \textit{ab initio} MD and saved for post-processing NAMD. 

\begin{align}
    \frac{\partial \ket{\psi(t)}}{\partial t} & = -iH\ket{\psi(t)} \\
    \ket{\psi(t)} & = \sum_i c_i(t)\ket{\phi_i(t)} 
\end{align}

In the practical implementation, following the previous P-matrix scheme~\cite{Zheng19p6174}, the time-dependent Hamiltonian ($H(t)$) within one MD step is obtained by a linear interpolation between the Hamiltonians of two successive MD steps: $H(T_m)$ and $H(T_{m+1})$, where $\Delta T=T_{m+1}-T_m$ is MD time step set as 2 fs. The Hamiltonians at MD time step $T_m$ and $T_{m+1}$ can be re-built by using the overlap matrix elements computed from the \textit{ab initio} MD and such two Hamiltonians share the same basis (index of $\mathbf{k}$ is not shown):

\begin{align}
     & \left<\phi_{i}(T_m)\middle|H(T_m+\Delta T)\middle|\phi_j(T_m)\right> \nonumber\\
     & = \sum_{k}\left<\phi_i(T_m)\middle|\phi_k(T_m+\Delta T)\right>\left<\phi_k(T_m+\Delta T)\middle|\phi_j(T_m)\right> \epsilon_k(T_m+\Delta T)
\end{align}

Here, it is necessary to note that the non-adiabatic couplings $V_{i\mathbf{k},j\mathbf{k'}}=\left\langle \phi_{i\mathbf{k}}(t) \middle| \frac{\partial \phi_{j\mathbf{k'}}(t)}{\partial t} \right\rangle\approx \left\langle \phi_{i\mathbf{k}}(t) \middle| \phi_{j\mathbf{k'}}(t+\Delta T) \right\rangle \frac{1}{\Delta T} -\delta_{i\mathbf{k},j\mathbf{k'}}$ are not used directly in our NAMD implementation, instead, the overlap matrix $S_{i\mathbf{k},j\mathbf{k'}}\equiv \left\langle \phi_{i\mathbf{k}}(t) \middle| \phi_{j\mathbf{k'}}(t+\Delta T) \right\rangle$ is used in equ. 19 to build Hamiltonians and evolve the wavefunctions. A few diagonalizations (usually less than 100) are performed within one MD step for $H(t)$ for a more accurate Hamiltonian. In this case, the eigen states of \textit{each} time step $dt$ ($dt$ is around 0.5 attosecond) are not chosen as the adiabatic states when evolving the above equ. 17. Instead, we choose the diagonalized eigen states within one MD step for the adiabatic states. At the start of each diagonalization, a basis change of adiabatic states is needed. As demonstrated in Ref.~\citenum{Zheng19p6174}, such implementation can significantly reduce the computational cost, particularly caused by diagonalizations. For our system, we find that 10 diagonalizations within one MD step already converge the calculation. 

Between the two diagonalizations, the wavefunciton and the Hamiltonian is written with the same basis (using the eigen states of the diagonalization as the basis). The wavefunction is evolved following the Schr\"{o}dinger's equation (equ. 17) with the small time step $dt$ as: $c_i(t+dt) = c_i(t) - i \sum_j\mathbf{H}_{ij}(t)c_j(t) dt - \frac{1}{2}\left[\mathbf{H}(t)\mathbf{H}(t)\right]_{ij} c_j(t)dt^2$. $c_i(t)$ is the coefficient of wavefunction expansion on the basis of adiabatic states (equ. 18). Here, it is necessary to note that at least the second order of $dt$ in the above expansion is needed to evolve the wavefunction correctly. Meanwhile, a converged $dt$ is needed. A $dt=0.5$ attosecond to show a converged carrier thermalization process for our system.

\begin{figure}
\centering
\includegraphics[width=0.8\columnwidth]{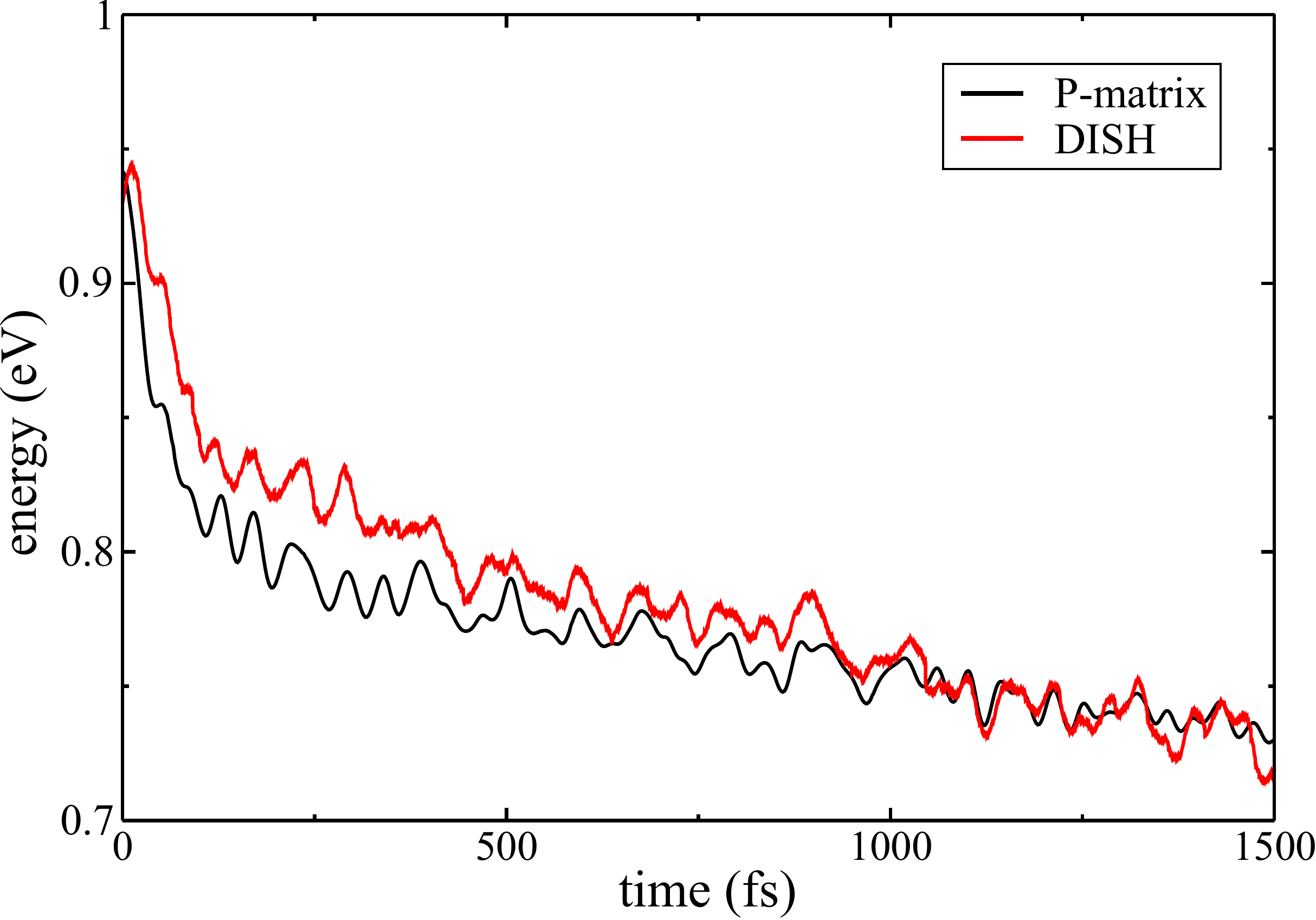}
\caption{A comparison between DISH and P-matrix for silicon 4$\times$4$\times$4 supercell with single $\Gamma$ $k$-point. The decoherence time for both calculations is set to 15 fs. 0 of $y$-axis is defined as the CBM of the system.}
\label{sfig6}
\end{figure}

Shown in SI Fig. 6 is the mean energy of the hot electron cooling between the P-matrix and DISH implementations performed for the 4$\times$4$\times$4 supercell with single $\Gamma$ $k$-point. These two methods show quite consistent cooling processes.

\section{6. Decoherence time}

DISH method can realize both the decoherence and detailed balance using the wavefunction collapsing method. Following the decoherence process proposed in Ref.~\citenum{Jaeger12p22A545}, the decoherence time enters our NAMD implementation as a constant time $\tau$ and a Poisson distribution of decoherence events with the expectation time-interval of $\tau$ is carried out. Here, for the simple bulk silicon system, we use a fixed decoherence time $\tau=15$ fs. This decoherence time is obtained following the simple method proposed in Ref.~\citenum{Akimov13p3857a} using the energy gap standard deviations. Meanwhile, Ref.~\citenum{Kang19p224303} also investigate a dynamically adjusted decoherence time between two eigen states for silicon quantum dots. A distribution of $\tau$ from 10 fs to 25 fs is obtained with its maximum distribution around 15 fs. SI Fig. 7 indicates the hot carrier thermalization process with different decoherence time (5 fs, 15 fs, and 25 fs) starting from the same initial state. Different decoherence-time cooling processes show relatively small differences.

\begin{figure}
\centering
\includegraphics[width=1\columnwidth]{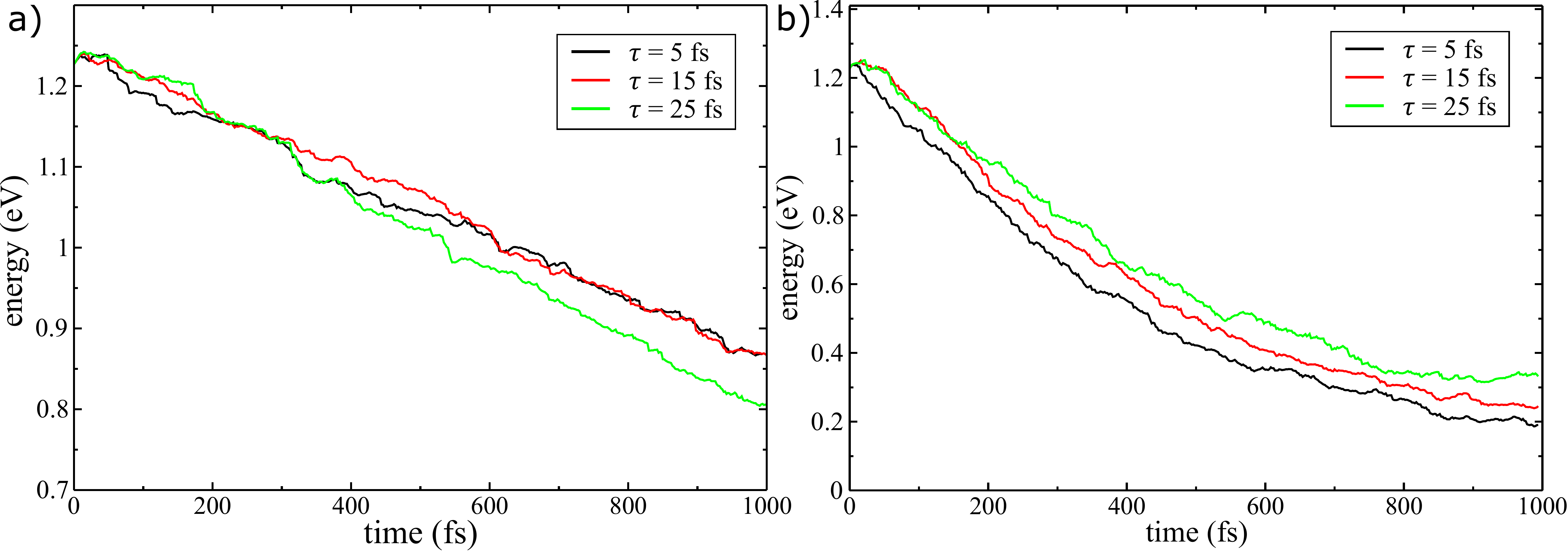}
\caption{A comparison of NAMD averaged energies starting from the same eigen state but with different decoherence time (5 fs, 15 fs, and 25 fs). a) thermalization process with the cross-$k$ transitions and b) thermalization process with the cross-$k$ transitions and the deformation potential (DP). 0 of $y$-axis is defined as the CBM of the system.}
\label{sfig7}
\end{figure}

\section{7. Orthonormalization of overlap matrix elements}

The overlap matrix is the key ingredient to build the Hamiltonian (equ. 19) and evolve the wavefunction. During the \textit{ab initio} MD, only the adiabatic states (i.e. eigen states $\phi_{i\mathbf{k}}$) within a finite energy window in which we are interested are selected to compute the overlap matrix elements. Thus, these states can not be used directly to span the Hilbert space of the wavefunction for the excited carriers ($\ket{\psi}=\sum_{i\mathbf{k}}c_{i\mathbf{k}}\ket{\phi_{i\mathbf{k}}}$). An orthonormalization is required to enforce the unitarity for the overlap matrix elements. Meanwhile, as illustrated in equ. 3 and 5 of main text, it is possible to have the accidental degeneracy between state $\epsilon_{i\mathbf{k}}(t)$ and state $\epsilon_{j\mathbf{k'}}(t+\Delta T)$ during the MD simulation, yielding an artificially large value. It is necessary to re-scale these spurious large cases during the NAMD simulation.

\subsection{7.1 Re-scale overlap matrix elements for accidental degenerate cases}

When $\mathbf{k}\neq\mathbf{k'}$, if two states have an accidental degeneracy ($\epsilon_{i\mathbf{k}}(t)\approx\epsilon_{j\mathbf{k'}}(t+\Delta T)$), it is necessary to re-scale these large overlap matrix elements. During the \textit{ab initio} MD, the matrix element $M_{i\mathbf{k},j\mathbf{k'}}\equiv\left<u_{i\mathbf{k}}(t)\middle|\Delta H\middle|u_{i\mathbf{k'}}(t+\Delta T)\right>$ will be computed and saved to files as well as eigen energies $\epsilon_{i\mathbf{k}}(t)$. Based on the equ. 3 in the main text, the overlap matrix element $S_{i\mathbf{k},j\mathbf{k'}}$ (used in equ. 19) is then computed with $M_{i\mathbf{k},j\mathbf{k'}}$ and re-scaled as:

\begin{align}
    S_{i\mathbf{k},j\mathbf{k'}} & = \left.\frac{1}{\sqrt{N_{\mathbf{k}}}} \sqrt{\frac{T_0}{T}} M_{i\mathbf{k},j\mathbf{k'}} {\rm sign}(\Delta \epsilon_{i\mathbf{k},j\mathbf{k'}}) \middle/ \left[\Delta \epsilon^2_{i\mathbf{k},j\mathbf{k'}} + \frac{2}{N_{\mathbf{k}}}\frac{T_0}{T}\left|M_{i\mathbf{k},j\mathbf{k'}}\right|^2\right]^{\frac{1}{2}}\right., \;\; \mathbf{k}\neq\mathbf{k'}
\end{align}

\noindent where $\Delta \epsilon_{i\mathbf{k},j\mathbf{k'}}$ is defined as $\epsilon_{j\mathbf{k'}}(t+\Delta T) - \epsilon_{i\mathbf{k}}(t)$.

When $\mathbf{k}=\mathbf{k'}$, the overlap $s_{i\mathbf{k},j\mathbf{k}}\equiv\left<u_{i\mathbf{k}}(t)\middle|u_{j\mathbf{k}}(t+\Delta T)\right>$ is computed and saved during the MD simulation following the previous implementations~\cite{Kang19p224303,Zheng19p6174}. Here, these diagonal-block parts should also subject to the $\frac{1}{\sqrt{N_{\mathbf{k}}}}\sqrt{\frac{T_0}{T}}$ factor to be used in DISH, except for the few elements near the diagonal positions whose overlaps $|s_{i\mathbf{k},j\mathbf{k}}|$ are close to 1 so as to $S_{i\mathbf{k},j\mathbf{k}}$ (the same state almost unchanged after $\Delta T$). To realize this property, the following re-scaling is applied as:

\begin{align}
    S_{i\mathbf{k},j\mathbf{k}} & = s_{i\mathbf{k},j\mathbf{k}} \left[1-\left(1-\frac{1}{\sqrt{N_{\mathbf{k}}}}\sqrt{\frac{T_0}{T}}\right)\left(1-\sin^4(\left|s_{i\mathbf{k},j\mathbf{k}}\right|^2\frac{\pi}{2})\right)\right], \;\; \mathbf{k}=\mathbf{k'}
\end{align}

\subsection{7.2 Orthonormalization procedure given the overlap matrix}

With the diagonal ($\mathbf{k}=\mathbf{k'}$) and off-diagonal ($\mathbf{k}\neq\mathbf{k'}$) blocks of $S_{i\mathbf{k},j\mathbf{k'}}$ ready, an orthonormalization procedure is carried out to unitarize the overlap matrix. First, a standard normalization process is performed. For a given state $i\mathbf{k}$, we have

\begin{align}
    \sum_{j\mathbf{k'}} \frac{\left|S_{i\mathbf{k},j\mathbf{k'}}\right|^2}{1+\alpha_{i\mathbf{k}}\left|S_{i\mathbf{k},j\mathbf{k'}}\right|^2} & = 1
\end{align}

$\alpha_{i\mathbf{k}}$ of each state $i\mathbf{k}$ can be solved using intersection method. After finding $\alpha_{i\mathbf{k}}$, the normalized $S'_{i\mathbf{k},j\mathbf{k'}}$ is defined as:

\begin{align}
    S'_{i\mathbf{k},j\mathbf{k'}} & = \frac{S_{i\mathbf{k},j\mathbf{k'}}}{\sqrt{1+\alpha_{i\mathbf{k}}\left|S_{i\mathbf{k},j\mathbf{k'}}\right|^2}}
\end{align}

Then, following the standard Lowdin symmetric orthogonalization procedure, $S'_{i\mathbf{k},j\mathbf{k'}}$ is further orthogonalized into $\widetilde{V}_{i\mathbf{k},j\mathbf{k'}}$.

\subsection{7.3 Orthonormalization procedure after adding deformation potential}

For the accidental degenerate cases, the magnitude of $S^{\rm DP}_{i\mathbf{k},j\mathbf{k'}}$ (following equ. 5 in main text but the overlap matrix is computed) is usually very large. The normalization of $S^{\rm tot}_{i\mathbf{k},j\mathbf{k'}}$ with the DP directly added to the overlap matrix tends to underestimate the magnitude of the cross-$k$ term. Therefore, the following procedure is carried out to add the DP induced non-adiabatic couplings.

Here, the overlap matrix $S^{\rm small}_{i\mathbf{k},j\mathbf{k'}}$ is built as:

\begin{align}
    S^{\rm small}_{i\mathbf{k},j\mathbf{k'}} & = 
    \begin{cases}
        S_{i\mathbf{k},j\mathbf{k}} & \text{diagonal block when $\mathbf{k}=\mathbf{k'}$} \\
        S^{\rm DP}_{i\mathbf{k},j\mathbf{k'}} & \text{off-diagnal block when $\mathbf{k}\neq\mathbf{k'}$} \\
    \end{cases}
\end{align}

\noindent where $S_{i\mathbf{k},j\mathbf{k}}$ and $S^{\rm DP}_{i\mathbf{k},j\mathbf{k'}}$ are all re-scaled following the above section 7.1 to remove spurious large values. The overlap matrix $S^{\rm small}_{i\mathbf{k},j\mathbf{k'}}$ is then orthonormalized to obtain $\widetilde{S}^{\rm small}_{i\mathbf{k},j\mathbf{k'}}$ following the above section 7.2. Now, we add the cross-$k$ overlap matrix elements. A new overlap matrix $S^{\rm new}_{i\mathbf{k},j\mathbf{k'}}$ is built as:

\begin{align}
    S^{\rm new}_{i\mathbf{k},j\mathbf{k'}} & = 
    \begin{cases}
        \widetilde{S}^{\rm small}_{i\mathbf{k},j\mathbf{k}} & \text{diaognal block when $\mathbf{k}=\mathbf{k'}$} \\
        S_{i\mathbf{k},j\mathbf{k'}} + \widetilde{S}^{\rm small}_{i\mathbf{k},j\mathbf{k'}} & \text{off-diagonal block when $\mathbf{k}\neq\mathbf{k'}$} \\
    \end{cases}
\end{align}

The $S^{\rm new}_{i\mathbf{k},j\mathbf{k'}}$ is further orthonormalized to obtain $\widetilde{S}^{\rm new}_{i\mathbf{k},j\mathbf{k}}$. The latter is then used in DISH algorithm.

\section{8. Convergence of number of trajectories}

The physical properties computed with surface hopping methods require a statistical average over the stochastic independent NAMD trajectories. SI Fig. 8 shows the calculated thermalization averaged energy carried out on different number of independent trajectories. Roughly around 100 trajectories are good enough to obtain the converged averaged energy. This number of trajectories is also used throughout this paper. 

\begin{figure}[h]
\centering
\includegraphics[width=1\columnwidth]{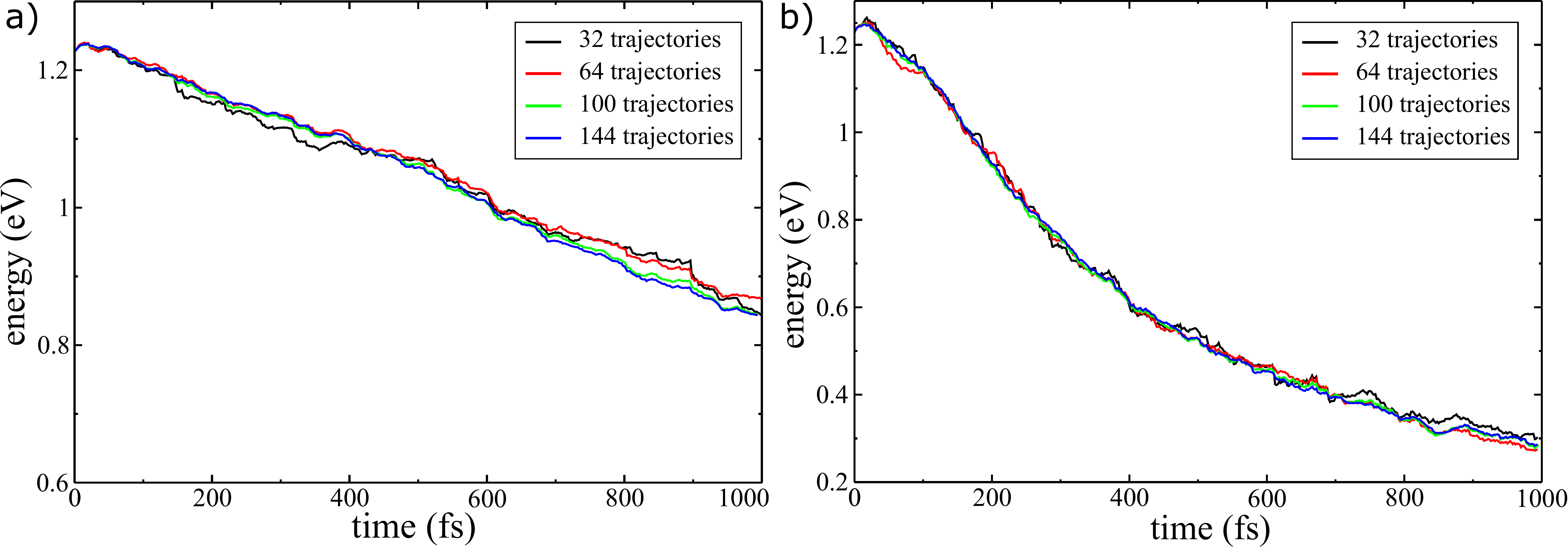}
\caption{A comparison of NAMD averaged energies starting from the same eigen state but with different number of trajectories for average. a) thermalization process with cross-$k$ transitions and b) thermalization process with cross-$k$ transitions and deformation potential (DP). 0 of $y$-axis is defined as the CBM of the system.}
\label{sfig8}
\end{figure}




\section{9. Convergence of $k$-point grids with DP}

After adding the deformation potential (DP), a convergence test of $k$-point grids is also carried out (SI Fig. 9). We find that the 2$\times$2$\times$2 $k$-point grid is still good.

\begin{figure}[h]
\centering
\includegraphics[width=0.6\columnwidth]{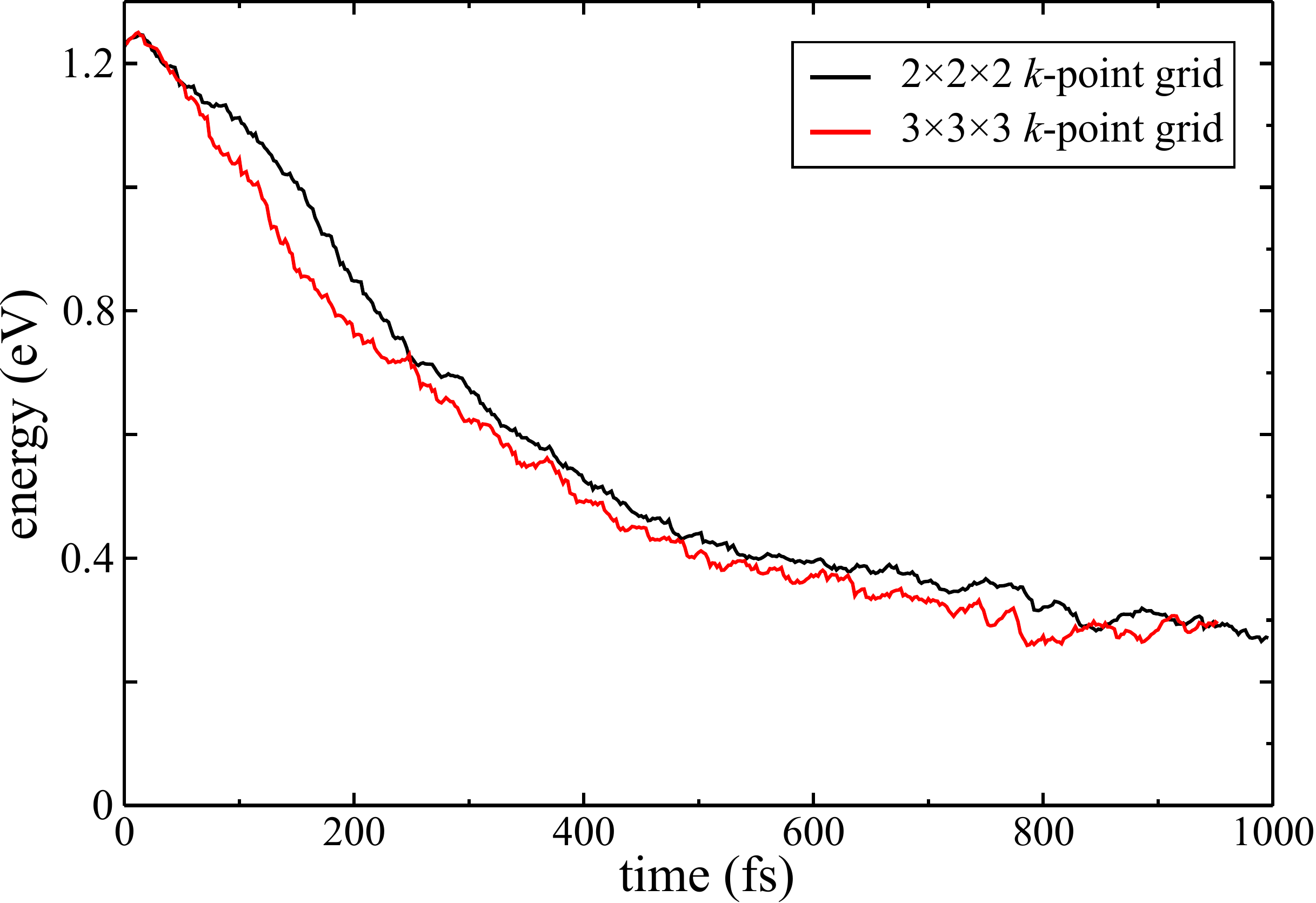}
\caption{The averaged energy of electron thermalization performed on the 2$\times$2$\times$2 and 3$\times$3$\times$3 $k$-point grids, respectively (for silicon 2$\times$2$\times$2 supercell). 0 of $y$-axis is defined as the CBM of the system.}
\label{sfig10}
\end{figure}

\section{10. Different deformation potential factor $E_1$}

Here, we find that the exact value of $E_1$ does not make a big difference to the overall energy-cooling profiles for NAMD. 

\begin{figure}[h]
\centering
\includegraphics[width=0.6\columnwidth]{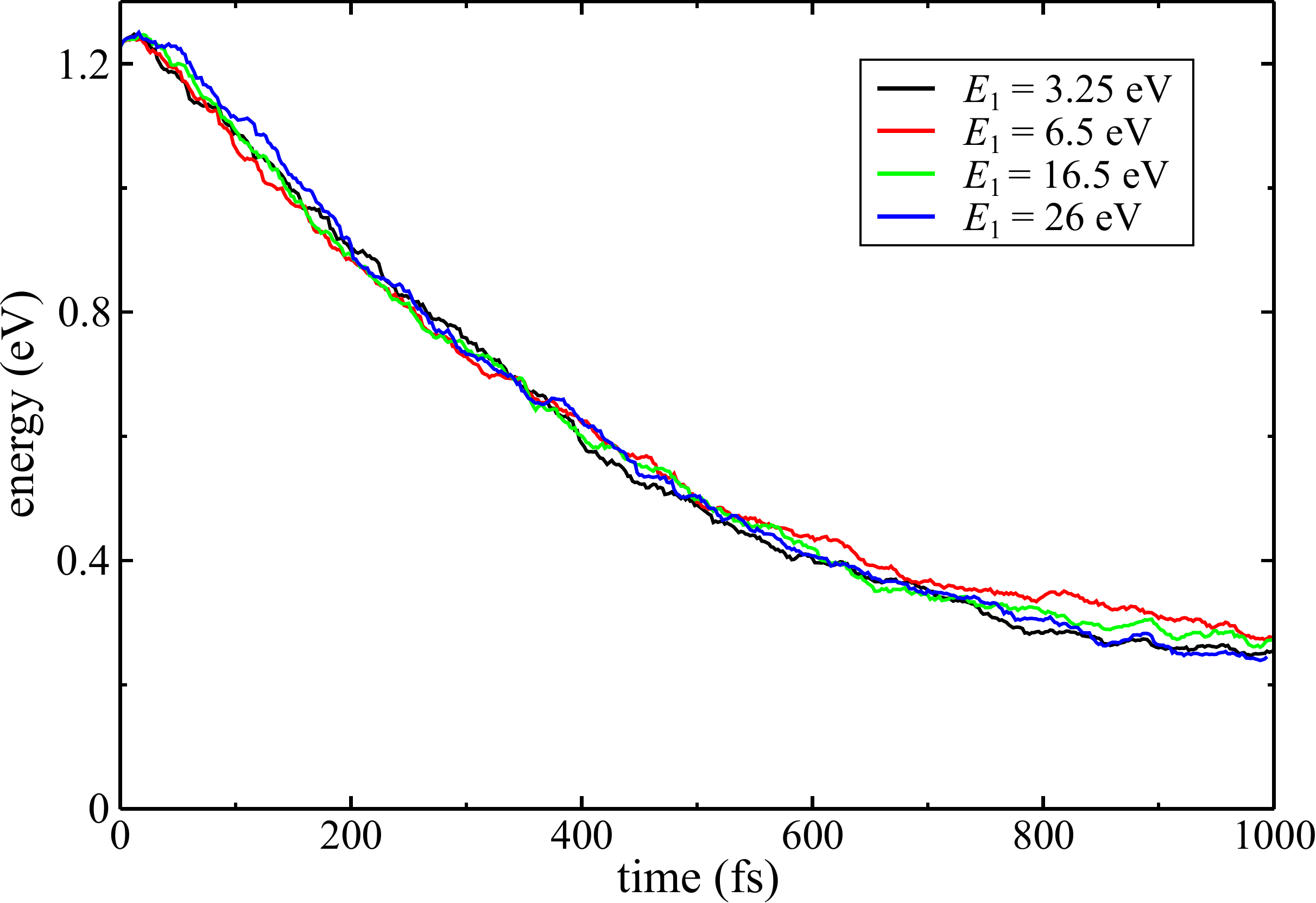}
\caption{The averaged energy of the electron thermalization performed on the 2$\times$2$\times$2 $k$-point grid for the 2$\times$2$\times$2 silicon supercell. 0 of $y$-axis is defined as the CBM of the system. $E_1 = 6.5$ eV is suggested by Ref.~\citenum{Bardeen50p72}.}
\label{sfig11}
\end{figure}

\section{11. An example illustration of $k$-resolved state population}

SI Fig. 11 presents three example NAMD trajectories to indicate the effects of the cross-$k$ transition and the DP. The left panel plots the average energy of \textit{one} NAMD trajectory. Shown in the right panel of SI Fig. 11 is the $k$-resolved wavefunction population as a function of time ($|c_{i\mathbf{k}}(t)|^2$). The eight $k$ points split the whole graph into eight horizontal stripes with the $y$-axis of each stripe indicates energy from 0 to 1.4 eV. Starting from around 1.2 eV in the 8th $k$ point ($k_8$), when no cross-$k$ transition is allowed (SI Fig. 11 top), the thermalization process is slowest with the population limited within one $k$ point. By turning on the cross-$k$ transitions (SI Fig. 11 middle), although the initial population is only occupied at one state, the wavefunction quickly spreads to the states belonging to all the other $k$ points within several femto-second. When the DP is added to strengthen the cross-$k$ transitions (SI Fig. 11 bottom), compared to the middle panel, we observe that the wavefunction populates more states, which is also owing to the enhanced cross-$k$ transitions.

\begin{figure}
\centering
\includegraphics[width=1.0\columnwidth]{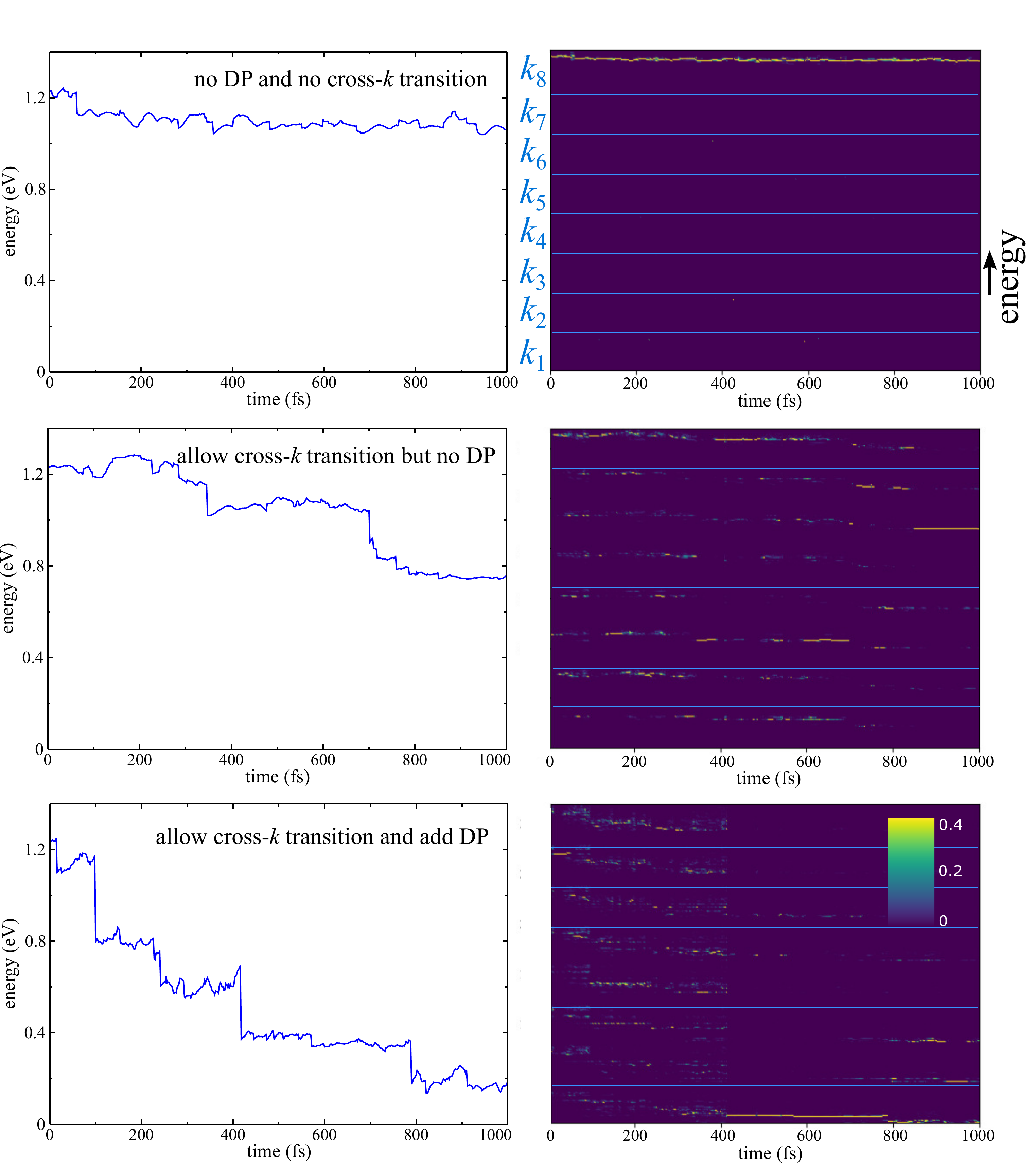}
\caption{Left panel: average energy for \textit{one} NAMD trajectory (no statistical average). 0 of $y$-axis is set as CBM of the system. Right panel: Wavefunction populations on the eigen states ($|c_{i\mathbf{k}}(t)|^2$) based on the trajectory shown in the left panel. Each graph is split into eight horizontal stripes with each stripe indicating states belonging to one $k$ point. The $y$-axis of each stripe has energy from 0 to 1.4 eV. Top two sub figures: a trajectory without cross-$k$ transitions; middle two sub figures: a trajectory by enabling the cross-$k$ transitions; bottom two sub figures: a trajectory by enabling the cross-$k$ transitions and the DP.}
\label{sfig12}
\end{figure}

\newpage


%